\documentclass[aps,prb,preprint,superscriptaddress,showpacs]{revtex4}
\usepackage{keyval}%
\usepackage{graphicx}
\usepackage{dcolumn}
\usepackage{bm}
\setkeys{Gin}{width=0.5\columnwidth,clip=true}
\newcommand{\ignore}[1]{}

\begin{document}
%\preprint{U8: 20842}

\title{Molecular dynamics simulation of the order-disorder phase transition in
solid NaNO$_2$}

\author{Wei-Guo Yin}
 %\altaffiliation[Also at ]{Physics Department, The Chinese University of Hong Kong.}%Lines break automatically or can be forced with \\
\email{wgyin@yahoo.com}
%\email{wyin@mail.unomaha.edu}
% \homepage{http://www.geocities.com/wgyin}
\author{Chun-Gang Duan}%
\author{W. N. Mei}%
\affiliation{%
Department of Physics, University of Nebraska, Omaha, NE 68182}%This line break forced with \\
\author{Jianjun Liu}
\affiliation{%
Department of Physics, University of Nebraska, Omaha, NE 68182}%This line break forced with \\
\affiliation{Department of Physics and Center for Electro-Optics,
University of Nebraska, Lincoln, NE 68588}%
\author{R. W. Smith}%
\affiliation{%
Department of Physics, University of Nebraska, Omaha, NE 68182}%This line break forced with \\
\author{J. R. Hardy}
\affiliation{Department of Physics and Center for Electro-Optics,
University of Nebraska, Lincoln, NE 68588}%


\date{\today}

\begin{abstract}
We present molecular dynamics simulations of solid NaNO$_2$ using
pair potentials with the rigid-ion model. The crystal potential
surface is calculated by using an \emph{a priori} method which
integrates the \emph{ab initio} calculations with the Gordon-Kim
electron gas theory. This approach is carefully examined by using
different population analysis methods and comparing the
intermolecular interactions resulting from this approach with
those from the \emph{ab initio} Hartree-Fock calculations.  Our
numerics shows that the ferroelectric-paraelectric phase
transition in solid NaNO$_2$ is triggered by rotation of the
nitrite ions around the crystallographical $c$ axis, in agreement
with recent X-ray experiments [Gohda \textit{et al.}, Phys.\ Rev.\
B \textbf{63}, 14101 (2000)]. The crystal-field effects on the
nitrite ion are also addressed. Remarkable internal
charge-transfer effect is found.
\end{abstract}

%\keywords{}

\maketitle

%\subsection{}
%\subsubsection{}

\section{Introduction}

\begin{figure}
\includegraphics*{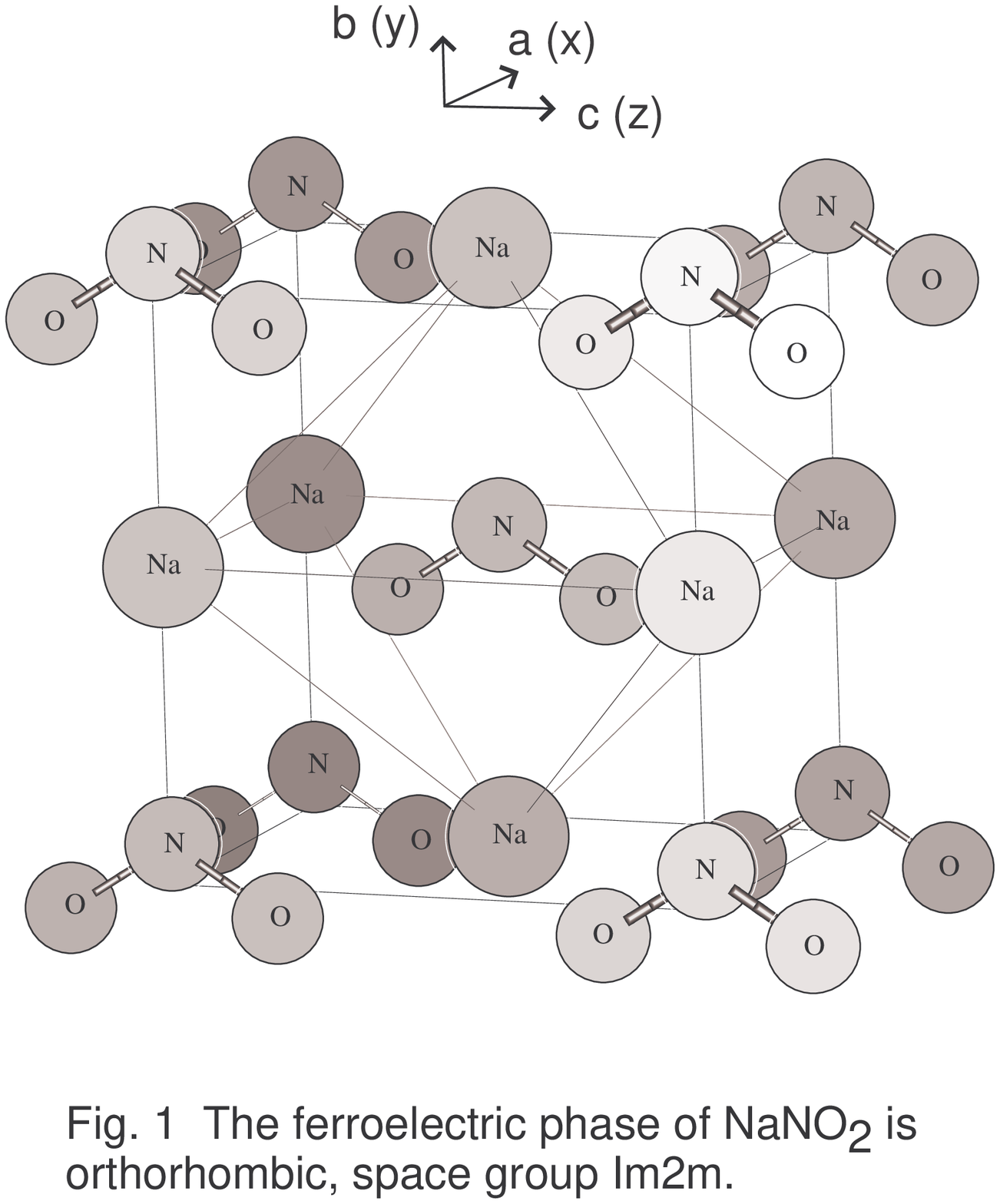}%
\caption{\label{fig:nano2}%
Crystal structure of NaNO$_2$ in the ferroelectric phase.}
\end{figure}

Sodium nitrite is a ferroelectric at room temperature. It has the
orthorhombic structure, space group $C_{2v}^{20}-Im2m$, with the
dipole vector of the V-shaped nitrite anions aligned parallel to
the crystallographic $b$ direction, as shown in
Fig.~\ref{fig:nano2}.
The ferroelectric-paraelectric phase transition takes place at
about $437$ K, where the high temperature phase is orthorhombic,
space group $D_{2h}^{25}-Immm$, with the dipoles disordered with
respect to the $b$ axis. In a narrow temperature range from
$435.5$ K to $437$ K, there exists an incommensurate
antiferroelectric phase. The melting temperature is $550$ K.
Distinguished from displacive ferroelectrics in which the
ferroelectric transition is driven by soft phonon modes, NaNO$_2$
offers a model system for research of the order-disorder
structural phase transition and any associated ferroelectric
instability. \cite{nano2:sawada,lines:glass,nano2:fokin}

Extensive experimental work on NaNO$_2$ has been devoted to
probing the mechanism of the NO$_2^{-}$ polarization reversal that
triggers the order-disorder transition. The majority of studies
support the $c$-axis rotation model, but there were also results
favoring the $a$-axis rotation model.\cite{nano2:gohda} Recently,
refined X-ray studies over a wide temperature range reinforced the
$c$-axis rotation model.\cite{nano2:gohda,nano2:komatsu}  On the
theoretical side, the microscopic model calculations done by
Ehrhardt and Michel supported the $c$-axis rotation
mechanism,\cite{nano2:michel} whereas mixed double rotations
around the $a$-axis and the $c$-axis was suggested by Kinase and
Takahashi.\cite{nano2:kinase} It has long been desirable to apply
computer molecular dynamics (MD) simulations to NaNO$_2$ in order
to achieve unambiguous understanding of the polarization reversal
mechanism. Earlier MD simulations with empirical Born-Mayer pair
potentials detected the $c$-axis rotation in
above-room-temperature NaNO$_2$.\cite
{nano2:klein,nano2:klein:london,nano2:klein_lynden-bell}
Unfortunately, the low-temperature structure produced by those
simulations was antiferroelectric and apparently disagreed with
the experimental observations.

Lu and Hardy pointed out that the overall phase behavior of
NaNO$_2$ could be simulated by using an \emph{a priori} approach
to construct the crystal potential surface
(PES).\cite{hardy:nano2}  The Lu-Hardy (LH) approach was
originally designed to deal with molecular crystals such as
K$_2$SeO$_4$, where exists a mix of bonding types, that is, the
intermolecular interactions are mostly ionic, but the constituent
atoms in a molecule (SeO$^{2-}_4$ in K$_2$SeO$_4$) bond
covalently. In the LH approach, the intra-molecule interactions
were treated by applying the \emph{ab initio} self-consistent
field method to the gas-phase molecules, while the intermolecular
pair potentials were computed within the Gordon-Kim (GK) electron
gas theory.\cite{hardy:k2seo4} The crux of their application of
the GK theory is how to partition the \emph{ab initio} molecular
charge density between the constituent atoms. Since there is no
unique way to separate the charge density of a highly covalently
bonded molecule, Lu and Hardy suggested equal separation in a
spirit similar to the Mulliken population analysis (MPA). By using
this atomic-level method, we could successfully describe the phase
transitions in fluoroperovskites,\cite{hardy:nacaf3} and ionic
crystals with polyatomic molecules including
SeO$^{2-}_4$,\cite{hardy:k2seo4_PRB} ClO$^-_4$,\cite {hardy:clo4}
SO$^{2-}_4$,\cite{hardy:k2so4}
SiO$^{4-}_4$,\cite{hardy:ca2sio4-sr2sio4} and NO$^-_3$.\cite
{hardy:kno3,hardy:rbno3-csno3,hardy:agno3}  Note that the MPA
happens to preserve the (zero) dipole moment of these molecules.

However, several problems appear when we moved on to deal with
NaNO$_2$ where the NO$^-_2$ radical has nonzero dipole moment and
stronger chemical bonding. First, it is well known that the MPA,
while certainly the most widely employed, is also somewhat
arbitrary and the most criticized.\cite{abinitio:hehre} In
particular, the MPA overestimates the dipole moment of the free
NO$_2^{-}$ ion by about $120\%$. Other difficulties involved the
free-ion approximation. Unlike in monatomic ionic crystals, there
may exist considerable \emph{internal} charge-transfer effects in
molecular ionic crystals. Electronic band structure calculations
\cite{nano2:ravindran} indicated that within a nitrite entity, the
nitrogen atom and two oxygen atoms bond covalently, leading to
high charge transferability between these constituent atoms.
Therefore, in solid NaNO$_2$ the NO$_2^{-}$ group will feel
different crystal-field environments as it rotates and responds by
redistributing the charge density among its three constituent
atoms.

Our goals in this paper are twofold. First, we show that our
atomistic level simulation methods involving pair potentials with
the rigid-ion model is capable of correctly describing the phase
behavior of NaNO$_2$. Second, we systematically examine the LH
approach to understand the reason why it works so well in
molecular ionic crystal systems by the following steps: (i) we
develop another population analysis method that preserves the
molecular dipole moment by directly fitting the \emph{ab initio}
charge density of a molecule; (ii) we carry out \emph{ab initio }
Hartree-Fock (HF) calculations of the intermolecular interactions
and find that the pair potentials from the rigid-ion model can
correctly reproduce the \emph{ab initio} results; (iii) we
investigate the crystal-field effects on the NO$_2^{-}$ ion by
embedding the ion (and its first shell of neighbors) in a lattice
of point charges and find a remarkable internal charge-transfer
effect. \cite{alkali_halide:fowler} Several MD simulations based
on these modifications of the LH approach are also performed. The
ferroelectric-paraelectric transition triggered by the $c$-axis
rotation of the nitrite ions is observed in all versions of the LH
approach. However, the transition temperatures predicted by these
simulations are lower than the experimental values. Furthermore,
the transition temperatures obtained from the original version are
higher than those predicted by modified versions and closer to the
experimental values. After careful examination, we notice that in
the LH approach, the NO$_2^-$ dipole moments were generally
overestimated by about 120\%, which reinforces the ground state.
This implies that the crystal structure of NaNO$_2$ is stabilized
by the anion polarization effects. Thus, we conclude polarization
effects are particularly important for the quantitative study of
NaNO$_2$.

This paper is organized as follows. Section II describes the
method we used to obtain the intermolecular interactions. Section
III analyzes the resulting intermolecular potentials in comparison
with those obtained from \emph{ab initio} calculations. Section IV
presents the results of our MD simulations. The crystal-field
effects on NO$_2^-$ are discussed in Section V. Concluding remarks
are made in Section VI.

\section{Intermolecular interactions}

Our MD simulation technique method originates from the GK model
for simple ionic crystals such as alkali halides, assuming that
(molecular) ions in a crystal environment may be described as free
ions. \cite
{gordon:kim:alkali-halide_alkaline-earth-dihalide,gordon:cohen:alkali-halide,boyer:nacl-kcl}
Then it was extended to deal with molecular ionic crystals like
K$_2$SeO$_4$, in which strong intramolecular covalency exists.
\cite{hardy:k2seo4,hardy:k2seo4_PRB} The main idea is that the molecular ion (SeO%
$_4^{-}$ in K$_2$SeO$_4$) is treated as a single entity, and
intramolecular and intermolecular interactions are treated
separately: first we perform \textit{ab initio} quantum chemistry
calculations for the whole molecular ion, to obtain the optimized
structure, the force constants, and the whole electron density
$\rho (\mathbf{r})$. The intramolecular interactions are described
by force constants within the harmonic approximation. As for the
intermolecular interactions, we have to carry out electron
population analysis to separate $\rho (\mathbf{r})$ onto each
individual atom in the molecular ion, then use the Gordon-Kim
electron gas model to calculate the intermolecular pair
potentials. This approach provides a parameter-free description
for the crystal potential-energy surfaces, which allow structural
relaxation, MD simulation, and lattice dynamics calculations.

In calculating the intermolecular forces, there are three major
approximations as discussed in the following:

(1) We assume that the geometries and electronic densities of the
separate ions remain unchanged once they have been obtained under
given circumstance, such as in the equilibrium state of the gas or
crystal phases. This approximation is the fundamental basis for
the GK electron gas theory. General speaking, we found that in an
ionic crystal there is no strong chemical bond between ions, hence
this approximation is reasonable.

(2) When dealing with the intermolecular interaction, we assume
that the charge density of a rigid ion can be separated into its
atomic constituents.

(3) We assume that the crystal potential energy is composed of the
intermolecular and intramolecular interaction, where the
intramolecular interaction is expressed in terms of force fields
and the intermolecular interaction is a sum of interatomic pair
potentials.

Atomistic level simulations utilizing pair potentials and the
rigid-ion model have great success in describing many ionic
systems. \cite{phillpot} We showed that this scheme can correctly describe the
phase transition behaviors of alkali halide fluoperovskites,
\cite{hardy:nacaf3} and molecular crystals with tetrahedral
\cite{hardy:k2seo4,hardy:k2seo4_PRB,hardy:clo4,hardy:k2so4,hardy:ca2sio4-sr2sio4}
and equilateral triangular \cite
{hardy:kno3,hardy:rbno3-csno3,hardy:agno3} radicals. However, for
NaNO$_2$ in which NO$_2^-$ has only a two fold symmetrical axis,
the results were less satisfactory. \cite{hardy:nano2} Note that
the mean HF polarizability of NO$_2^{-}$, 14.156 (atomic units),
calculated with the D95* basis, \cite{gaussian} is much higher
than that of Na$^+$, 0.343 (atomic units), calculated with the
6-31* basis. Therefore, in solid NaNO$_2$, the rotation of
NO$_2^{-}$ in the crystal field will induce charge redistribution
within the molecule. Hence this dynamic effect may invalidate the
rigid ion approximation. In this paper we shall perform HF
calculations for various geometries to verify this scenario.

\subsection{Pairwise additive approximation}

In the GK model, we evaluate the interaction between two molecules
based on the electron density, \cite{gordon:kim} which is
approximated as the sum of component densities taken from HF
calculations. That is, if $\rho _A$ and $\rho _B$ are the
component densities, then the total density is $\rho _{AB}^{}=\rho
_A^{}+\rho _B^{}$. Whereas, interaction potential is computed as
the sum of four terms: Coulombic, kinetic, exchange, and
correlation energies which are expressed in terms of the charge
densities.

Therefore, suppose the $A$ and $B$ molecules are made up of $M$
and $N$ atoms, respectively, then the Coulombic interaction
between them is

\begin{eqnarray}
V_\mathrm{C} &=&\int \int d\mathbf{r}_1d\mathbf{r}_2\frac{\rho _A^{}(\mathbf{r}%
_1)\rho _B^{}(\mathbf{r}_2)}{|\mathbf{r}_1-\mathbf{r}_2|}%
-\sum_{i=1}^MZ_{A,i}\int d\mathbf{r}_2\frac{\rho _B^{}(\mathbf{r}_2)}{|%
\mathbf{r}_2-\mathbf{R}_{A,i}|}  \nonumber \\
&&-\sum_{j=1}^NZ_{B,j}\int d\mathbf{r}_1\frac{\rho _A^{}(\mathbf{r}_1)}{|%
\mathbf{r}_1-\mathbf{R}_{B,j}|}+\sum_{i=1}^M\sum_{j=1}^N\frac{Z_{A,i}Z_{B,j}%
}{|\mathbf{R}_{A,i}-\mathbf{R}_{B,j}|},  \label{Coul}
\end{eqnarray}
where $Z_{A,i}$, $Z_{B,j}$, $\mathbf{R}_{A,i}$ and
$\mathbf{R}_{B,j}$ are the nuclear charges and coordinations of
the $i$-th atom in the $A$ molecule and $j$-th atom in the $B$
molecule, respectively. This potential energy can be split into
two parts: first the long-range part,
\begin{equation}
V_\mathrm{C}^l=\sum_{i=1}^M\sum_{j=1}^N\frac{[Z_{A,i}-\int \rho _A^{}(\mathbf{r}_1)d%
\mathbf{r}_1][Z_{B,j}-\int \rho _B^{}(\mathbf{r}_2)d\mathbf{r}_2]}{|\mathbf{R%
}_{A,i}-\mathbf{R}_{B,j}|}, \label{cl}
\end{equation}
and the short-range part
\begin{equation}
V_\mathrm{C}^s=V_\mathrm{C}-V_\mathrm{C}^l.  \label{cs}
\end{equation}
Eq. (\ref{cl}) is essentially the electrostatic interaction energy
when the charge densities of the molecules are distributed as
point charges on the constituent atoms, which is known as the
Madelung potential energy.

The non-Coulombic energy terms are expressed in the uniform electron gas
formula,
\begin{equation}
V_i=\int d\mathbf{r[}\rho _{AB}^{}(\mathbf{r})\epsilon_i(\rho
_{AB}^{})-\rho _A^{}(\mathbf{r})\epsilon _{i}(\rho _A^{})-\rho
_B^{}(\mathbf{r})\epsilon _{i}(\rho _B^{})], \label{kxc}
\end{equation}
where $\epsilon_i(\rho )$ is one of the energy functionals for the
kinetic, exchange, and correlation interactions. \cite{gordon:kim}
Note that Eq. (\ref{kxc}) is not composed of pair potentials. In
order to obtain the effective pairwise potentials, we approximate
Eq. (\ref{kxc}) using

\begin{equation}
V_i\simeq \sum_{m\in A}\sum_{n\in B}\int d\mathbf{r[}\rho
_{mn}^{}(\mathbf{r})\epsilon _{i}^{}(\rho _{mn}^{})-\rho _m^{}(%
\mathbf{r})\epsilon _{i}^{}(\rho _m^{})-\rho _n^{}(\mathbf{r}%
)\epsilon _{i}^{}(\rho _n^{})],  \label{pair}
\end{equation}
where $\rho _{mn}^{}=\rho _m^{}+\rho _n^{}$. $\rho _m^{}$ and
$\rho _n^{}$ are the charge densities of individual atoms in the
$A$ and $B$ molecules, respectively, which are obtained by a
population analysis as described in the next subsection.

Even though the non-Coulombic forces as determined by Eq.
(\ref{kxc}) are not strictly additive, the above approximation
appears to be adequate except at very short distances. As pointed
out by Waldman and Gordon, \cite {gordon:waldman:shell} the main
reason as to why this approximation is valid is because the
Coulombic force, the largest contribution to the potentials, is
additive. Based on our calculations, we find additivity of $V_i$
holds only to within about 50\%; however, the overlap contribution
to the electrostatic energy dominates $V_i$ and renders additivity
to within 10\%. One final remark is in order, for the sake of
simplifying the two-electron integral in Eq. (\ref {Coul}), the
charge densities, $\rho _m^{}$ and $\rho _n^{}$, are taken as its
spherical average. As a result, the Coulombic interaction is not
exactly evaluated. Nevertheless, as we shall show in
Figs.~\ref{fig:na-no2} and \ref{fig:no2-no2}, this error is
compensated by those due to the pairwise additive approximation.

To summarize this subsection, we demonstrated that it is possible
to analytically express the intermolecular potentials
$V_\mathrm{C}^l+V_\mathrm{C}^s+V_i$ using Eqs. (\ref{cl}),
(\ref{cs}) and (\ref {pair}) once the charge density of each
individual atom is obtained by an electronic population analysis.
In the next subsection, we shall present further analysis on the
charge density.

\subsection{Electronic Population Analysis}

In this subsection, we discuss the ways to separate the electron density $%
\rho (\mathbf{r})$ of a molecule into its atomic constituents.
Suppose the molecule consists of $M$ atoms, then the wave function of the molecule $\psi (%
\mathbf{r})$ can be written as a linear superposition of atomic
wave functions ${\varphi (\mathbf{r}-\mathbf{R}_i)}${,}
$i=1,2,\ldots ,M$, centered at each atom,

\begin{equation}
\psi (\mathbf{r})=\sum_{i=1}^M{\varphi (\mathbf{r}-\mathbf{R}_i).}
\end{equation}
In turn, the atomic wave functions ${\varphi (\mathbf{r}-\mathbf{R}_i)}$ can
be written as a linear superposition of the basis functions $\chi _l^{}$

\begin{equation}
{\varphi (\mathbf{r}-\mathbf{R}_i)=}\sum_lc_{il}^{}\chi _l^{}(\mathbf{r-}{%
\mathbf{R}_i}).
\end{equation}
where \{$\chi _l^{}(\mathbf{r-}{\mathbf{R}_i})$\} are usually the
gaussian basis functions, and the coefficients $c_{il}^{}$ can be
obtained from the variational method.

Then the electronic density of the molecule is,
\begin{equation}
\rho (\mathbf{r})=|\psi (\mathbf{r})|^2=\sum_{ijkl}d_{ik,jl}^{}\chi _k^{}(%
\mathbf{r-}{\mathbf{R}_i})\chi _l^{}(\mathbf{r-}\mathbf{R}_j),
\label{density}
\end{equation}
where $d_{ik,jl}^{}=2c_{ik}^{}c_{jl}^{}$, which can be divided
into two parts, namely the \textit{net} ($i=j$) and
\textit{overlap} ($i\neq j$) populations. The latter cannot be
ignored in the presence of strong intramolecular covalency.
Therefore, separating $\rho (\mathbf{r})$ into its atomic
constituents is to split the overlap population. However, the way
to achieve that is not unique. For example, we can introduce a set
of weights $w_{ijkl}$ due to different criteria such that
\begin{eqnarray}
\widetilde{d}_{ik,ik}^{} &=&d_{ik,ik}^{}+\sum_{j\neq
i,l}w_{ijkl}d_{ik,jl}^{}\int \chi _k^{}(\mathbf{r-}{\mathbf{R}_i})\chi _l^{}(%
\mathbf{r-}\mathbf{R}_j)d\mathbf{r},  \nonumber \\
\widetilde{d}_{jl,jl}^{} &=&d_{jl,jl}^{}+\sum_{i\neq
j,k}(1-w_{ijkl})d_{ik,jl}^{}\int \chi _k^{}(\mathbf{r-}{\mathbf{R}_i})\chi
_l^{}(\mathbf{r-}\mathbf{R}_j)d\mathbf{r},  \label{w} \\
\widetilde{d}_{ik,il}^{} &=&d_{ik,il}^{},  \nonumber
\end{eqnarray}
then we can rewrite Eq. (\ref{density}) as following
\begin{equation}
\rho (\mathbf{r})\simeq \sum_i\rho _i(\mathbf{r})=\sum_{ikl}\widetilde{d}%
_{ik,jl}^{}\chi _k^{}(\mathbf{r-}{\mathbf{R}_i})\chi _l^{}(\mathbf{r-}%
\mathbf{R}_i),  \label{rho}
\end{equation}
where $\rho _i(\mathbf{r})$ is the atomic density of atom $i$.

\begin{table}%[H] add [H] placement to break table across pages
\caption{\label{table:multipole} Electronic multipole moments of
molecule ($AB_n$) calculated from the Mulliken population
analysis. The {\it ab initio} values are shown in parentheses. All
quantities are in atomic units.}
\begin{ruledtabular}
\begin{tabular}{lcccc}
$AB_n$\footnotemark[1] & $\mu_z$\footnotemark[2] & $\vartheta_{zz}$ & $\Omega_{zzz}$ & $\Phi_{zzzz}$ \\
\hline
CN$^-_{}$       &    0.14 (0.17) &    -27.35 (-29.41)    & 7.99 (10.68) & -143 (-166)   \\
NO$^-_2$     &    0.57 (0.26) &    -19.64 (-21.64)    & -2.60 (-5.87) &   -60 (-76)   \\
NO$^-_3$     &       0   (0)   &   -15 (-16)    &   0   (0)   &     -38 (-41)   \\
CO$^{2-}_3$  &       0   (0)   &   -17 (-17)    &   0   (0)   &     -48 (-48)   \\
ClO$^-_4$    &       0   (0)   &   -111 (-111)   &   0   (0)   &    -548 (-548)   \\
SO$^{2-}_4$  &       0   (0)   &   -119 (-120)   &   0   (0)   &    -620 (-627)   \\
SeO$^{2-}_4$ &       0   (0)   &   -143 (-143)   &   0   (0)   &    -823 (-816)   \\
SiO$^{4-}_4$ &       0   (0)   &   -157 (-158)   &   0   (0)   &   -1011 (-1019)   \\
ScF$^{3-}_6$ &       0   (0)   &   -319 (-318)   &   0   (0)   &   -5112 (-5071)   \\
SnCl$^{2-}_6$&      0   (0)   &   -845 (-844)   &   0   (0)   &  -19834 (-19668)   \\
\end{tabular}
\end{ruledtabular}
\footnotetext[1]{In the HF calculations, basis set D95* were used
for $AB$ and $AB_2$, 6-31G* for $AB_3$ and $AB_4$, and 3-21G* for $AB_6$.}%
\footnotetext[2]{The electrostatic moments $\mu$ (dipole),
$\vartheta$ (quadrupole), $\Omega$ (octapole), and $\Phi$
(hexadecapole) refer to the center of mass of the molecule with
the standard orientation defined in GAUSSIAN 98
(Ref.~\onlinecite{gaussian}).}
\end{table}

In our previous studies, the overlap electronic density is equally
separated, i.e., $w_{ijkl} = 1/2$ in Eq. (\ref{w}), similar to the
Mulliken population analysis (MPA). \cite
{hardy:nano2,hardy:k2seo4,hardy:kno3,hardy:ca2sio4-sr2sio4,hardy:nitrite}
In Table~\ref{table:multipole}, we present the electronic
multipole moments of SnCl$^{2-}_6$, ScF$^{3-}_6$, SiO$^{4-}_4$,
SeO$^{2-}_4$, SO$^{2-}_4$, ClO$^-_4$, CO$^{2-}_3$, and NO$^-_3$
calculated from using the MPA, and compare them with the \emph{ab
initio} values. We note that for these symmetrical molecules, the
MPA preserves total charge and zero dipole moment. However, for a
molecular ion like V-shaped NO$^-_2$ or linear CN$^-_{}$, the
Mulliken population seems inadequate. Given total charge and the
dipole moment, 0.26 a.u., of $NO^-_2$ obtained from the \textit{ab
initio} calculations (Table~\ref{table:multipole}), a population
analysis which preserves these vales would give rise to -0.092$e$
on N and -0.454$e$ on O. Whereas, using the MPA ($w_{ijkl} =
1/2$), the charges on the N and O atoms are 0.1624$e$ and
-0.5812$e$, respectively, which renders the dipole moment 0.57
atomic unit, overestimated by 120\%.

Therefore, it is desirable to determine $w_{ijkl}$ in such a way
that the calculated multipole moments of the molecule are
consistent with the \emph{ab initio} values. One possible way is
to evaluate $w_{ijkl}$ by fitting the \textit{ab initio} charge
density, as shown in Eqs. (\ref{w}) and (\ref{rho}), with the
values of multipole moments as constraints. An alternative way is
to directly fit the charge density, Eq. (\ref{rho}), with
$\widetilde{d}_{ik,jl}^{}$ being the parameters and
$\chi _k^{}(\mathbf{r-}{\mathbf{R}_i})\chi _l^{}(\mathbf{r-}%
\mathbf{R}_i)$ being the dependent variables. To simplify the
computation, only the radial parts of
$\chi _k^{}(\mathbf{r-}{\mathbf{R}_i})\chi _l^{}(\mathbf{r-}%
\mathbf{R}_i)$ were kept. This fitting population analysis (FPA)
is similar to that proposed by Parker and his co-workers as an
alternative implementation of the GK model. \cite{hf:parker}

%\input{table-frequency}
%
%To further elucidate the influence of strong covalency in
%NO$^-_2$, we preform a bond-length analysis. In
%Table~\ref{table:frequency}, we present the bond length $R_{AB}$
%of molecule $AB_n$ at its optimized molecular geometry and compare
%it with $r_A+r_B$ where $r_A$ and $r_B$ are the radii of the $A$
%and $B$ atoms with full ionicity, respectively.
%
%Together, we also list their $\omega_\textrm{min}$ (Kelvin), the
%lowest phonon frequencies of internal motion, and $T^*$, the phase
%transition temperatures in certain compounds.
%
%It is a common practice that we use $R_{AB}/(r_A+r_B)$ as a
%measure for covalent bonding, that it, the stronger the bond, the
%smaller the ratio. Among the ionic molecules listed in the table,
%$R_{AB}/(r_A+r_B)$ for NO$^-_2$ is the smallest. This ratio should
%be smaller for NO$^-_2$ than for NO$^-_3$ because the N atom in
%NO$^-_2$ has less ionic character than in NO$^-_3$. Also in
%Table~\ref{table:frequency}, Clearly, the internal motion of
%NO$^-_2$ is hardly excited by thermal energy than that of
%NO$^-_3$, reflecting that the covalent bonding in NO$^-_2$ is very
%strong. The goodness of partitioning the charge density of a
%molecule is determined by the covalency of the chemical bonds in
%the molecule. The stronger the covalent bonding, the worse the
%partitioning.

\subsection{Free ion and intramolecular interactions}

%The first requirement is an accurate representation of the ions.

GAUSSIAN 98 program package \cite{gaussian} is employed to obtain
the self-consistent solutions to the Hartree-Fock-Roothan
equations. The atomic orbital basis sets used are a double-zeta
basis with polarization functions (D95*) for the nitrogen and
oxygen atoms. It was shown that this choice of basis set could
give a good description of free NO$_2^{-}$.
\cite{no2-h2o:banerjee} As for the sodium atoms, we used both the
standard 6-31G* basis and the Slater-type orbitals for Na$^{+}$
taken from the Clementi and Roetti table, \cite{clementi} it
turned out the difference between them is small.

We first consider the free ion case and will discuss the
in-crystal ions in Section IV. The molecular geometry of
NO$_2^{-}$ can be described by internal coordinates $(
R_\mathrm{NO_1}$, $R_\mathrm{NO_2}, \theta_\mathrm{ONO})$ where
$R_\mathrm{NO_1}$ and $R_\mathrm{NO_2}$ are the lengths of the two
N-O bonds, respectively, and $\theta_\mathrm{ONO}$ is the angle
formed by the two N-O bonds. The minimum energy geometry of
NO$_2^{-}$ using our basis
yields $R^0_{\mathrm{NO}}=1.233$ \AA\ and $\theta^0_{%
\mathrm{ONO}}=116.6{{}^{\circ }}$. These structural parameters are
comparable to the experimental geometry of NO$_2^{-}$ in the
ferroelectric phase of NaNO$_2$, \cite{nano2:kay} in which
$R^0_{\mathrm{NO}}=1.236$ \AA\ and $\theta^0
_{\mathrm{ONO}}=115.4{{}^{\circ }}$. The internal vibration
frequencies of the NO$_2^{-}$ group are then calculated at the
optimized geometry. Note that the lowest vibration frequency (1192
K) of NO$^-_2$ is considerably higher than the highest libration
frequency (318 K) obtained from Raman spectroscopy
\cite{nano2:hartwig} as well as the order-disorder transition
temperature (437 K). Therefore, it is justified to treat the
internal motion of the nuclei in the NO$_2^{-}$ group within the
harmonic approximation, or even as a rigid rotor. To further
support this argument, we found that only small changes in the
geometry of NO$_2^{-}$ were observed in the high temperature
paraelectric phase ($R^0_{\mathrm{NO}}=1.239$ \AA\ and $\theta^0
_{\mathrm{ONO}}=114.5^\circ$). \cite{nano2:komatsu}

The intramolecular interactions are then represented by the
following force field obtained from frequency analysis in GAUSSIAN
98: \cite{gaussian}
\begin{eqnarray}
U=&& U^{}_0 +
\frac{1}{2}k^{}_\mathrm{NO}[(R^{}_\mathrm{NO_1}-R^0_\mathrm{NO})^2+
(R^{}_\mathrm{NO_2}-R^0_\mathrm{NO})^2]+
\frac{1}{2}k^{}_\mathrm{ONO}(\theta^{}_\mathrm{ONO}-\theta^0_\mathrm{ONO})^2
+ \nonumber \\
&&
k^{}_{2}(R^{}_\mathrm{NO_1}-R^0_\mathrm{NO})(R^{}_\mathrm{NO_2}-R^0_\mathrm{NO})
+k^{}_{3}(R^{}_\mathrm{NO_1}+R^{}_\mathrm{NO_2}-2R^0_\mathrm{NO})(\theta^{}_\mathrm{ONO}-\theta^0_\mathrm{ONO})
\end{eqnarray}
where $U^{}_0=-204.121$, $R^0_\mathrm{NO}=2.330$,
$\theta^0_\mathrm{ONO}=2.036$, $k^{}_\mathrm{NO}=0.728$,
$k^{}_\mathrm{ONO}=0.677$, $k^{}_2=0.174$, $k^{}_3=0.066$ in
atomic units.

Note that the polarizability of NO$_2^{-}$ at its optimized
geometry is highly anisotropic, that is, with $\alpha
_{xx}=7.820$, $\alpha _{yy}=10.823$, $\alpha _{zz}=23.825$ in
atomic units (see Fig.~\ref{fig:nano2} for coordinate convention).
Thus, one would expect this polarization to seriously affect the
intermolecular pair potentials, and thus renders the rigid-ion
approximation in question. Nevertheless, after extensive \emph{ab
initio} HF calculations with various configurations, which will be
presented in the next subsection, we find that the intermolecular
potentials for the NO$_2^{-}$:Na$^{+}$ and NO$_2^{-}$:NO$_2^{-}$
dimers can still be correctly reproduced.

\section{intermolecular potentials}

We scan the potential energy surface of NO$%
_2^{-}$:Na$^{+}$ and NO$_2^{-}$:NO$_2^{-}$ obtained from the
Hartree-Fock method. In these calculations, the geometrical
variables of NO$_2^{-}$ are frozen at their equilibrium values,
since we showed previously that the NO$_2^{-}$ group in NaNO$_2$
could be treated as a rigid rotor. However in the \textit{ab
initio} HF calculations, the electronic structure is allowed to
vary in order to minimize the total energy, thus the electronic
polarization effects are included.

In our calculations of intermolecular interactions, one NO$_2^{-}$
is fixed with its center of mass being the origin of the
coordinate system, the dipole vector pointing to the $y$ axis and
the O-O line being aligned parallel to the $z$ axis (see
Fig.~\ref{fig:nano2}). Then, Cartesian coordinates ($x$, $y$, $z$)
are the position of Na$^{+}$ or the center of mass of another
NO$^-_2$. To describe the orientation of the NO$_2^{-}$ molecule
at ($x$, $y$, $z$), we use the angles $\alpha $ and $\beta $ of
the dipole moment vector of the NO$_2^{-}$ molecule and an angle
$\gamma $ involving the rotation of NO$_2^{-}$ around its dipole
vector.

In order to study the effects of the different partition schemes,
mentioned in Section IIB, on the pair potentials, we performed
three different calculations, namely (i) MPA with pair potentials
[Eq. (\ref{pair})], (ii) FPA with pair potentials, (iii) FPA with
non-pair potentials as shown in Eq. (\ref {kxc}). We shall refer
them as models I, II, and III, respectively, from now on. Recall
that model I overestimates the dipole moment of NO$_2^-$ as
mentioned in Section IIB. Furthermore, we shall show in the
following that the electronic polarization effect could be
revealed from examining the differences between model I and II,
while the errors due to the pairwise additive approximation
employed by the GK model could be analyzed from the differences
between model II and III.

%\subsection{The NO$_2^{-}$:Na$^{+}$ intermolecular potential}

\begin{figure}
\includegraphics*{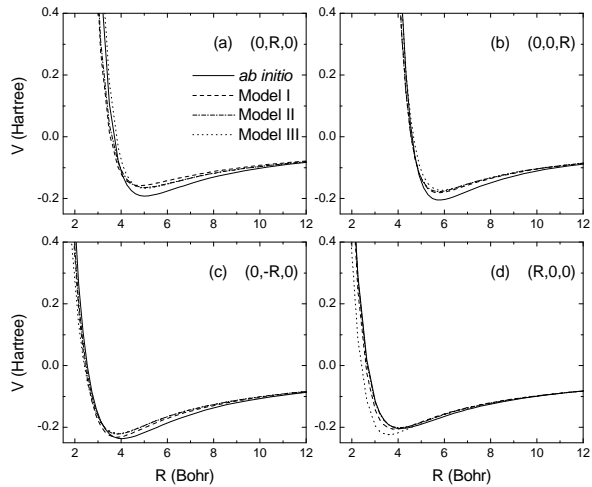}%
\caption{\label{fig:na-no2}%
NO$_2^{-}$-Na$^{+}$ intermolecular potential energy curves as a
function of $R$ for various configurations: $(0,R,0)$, $(R,0,0)$,
$(0,-R,0)$, and $(0,0,R)$, where $(x,y,z)$ is the location of
Na$^+$. Different lines represent the \textit{ab initio} HF model
(solid), model I (dashed), model II (dotted), and model III
(dashed and dotted). }
\end{figure}

\begin{figure}
\includegraphics*{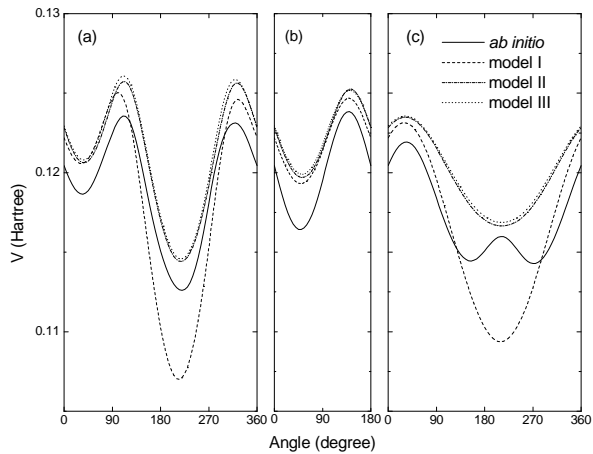}%
\caption{\label{fig:no2-no2}%
NO$_2^{-}$-NO$_2^{-}$ intermolecular potential energy curves as a
function of rotation angle. The NO$_2^{-}$ molecules are initially
in parallel alignment at separation ($1.82$ \AA , $2.83$ \AA ,
$2.69$ \AA ) and then one of them rotates around one of the (a)
$x$, (b) $y$ and (c) $z$ axes through its center of mass. }
\end{figure}

In Figs.~\ref{fig:na-no2} and \ref{fig:no2-no2}, we compare the
pair potentials obtained from these models and those from the
\textit{ab initio} HF calculations. In Fig.~\ref{fig:na-no2}
we show the results for the NO$_2^{-}$:Na$^+_{}$ dimer: we find
that in both models I and II, the overall shapes of the GK
potentials as a function of molecular separation are in agreement
with the \textit{ab initio} results, particularly in
Fig.~\ref{fig:na-no2}(d). On the other hand, the electronic
polarization effect also manifests itself in Fig.~\ref{fig:na-no2}
based on the following two observations. First, notice
Figs.~\ref{fig:na-no2}(b) and \ref{fig:na-no2}(d), models I and II
fit the \textit{ab initio} results through the entire range in
$(R,0,0)$, whereas in $(0,0,R)$ models I and II overestimate the
potentials when $5<R<8$. We attribute that to
the anisotropic polarizability of NO$_2^{-}$ ($%
\alpha _{xx}<\alpha _{yy}<\alpha _{zz}$), thus the electronic
cloud of NO$_2^{-}$ is most unlikely to be polarized along the
$(R,0,0)$ direction, along which the polarization is the weakest,
rather than the $(0,0,R)$ direction. Second, in
Fig.~\ref{fig:na-no2}(a), $(0,R,0)$, the minimum potential energy
in model II is closer to the \textit{ab initio} values than model
I, whereas in $(0,-R,0)$, where Na$^+$ is located below NO$_2^-$,
we found the situation reversed. To understand this, we observe
that in the $(0,R,0)$ configuration, Na$^+$ is closer to N than O,
thus in the \textit{ab initio} calculations the electrons were
attracted toward the N atom and thus led to a smaller dipole
moment. Therefore, model II, which produced smaller dipole moment
than model I, tends to get closer to the HF results. Obviously,
the results will be opposite if the situation is reversed.

Among the four different configurations, Figs.~\ref{fig:na-no2}(a)
through \ref{fig:na-no2}(d), and methods except model III, the
lowest NO$_2^{-}$-Na$^+_{}$ potential energy takes place at
$(0,-R,0)$, which will be shown later to give rise to the lowest
energy structure of NaNO$_2$.
%, whereas the minimum energies obtained from model III
%occurs at the $(R,0,0)$ configuration.
Both models II and III use FPA, the only difference between them
is that model II employs the pair potentials. It seems that in
model II the errors caused by the pairwise additive approximation
is compensated by the errors due to FPA.

%\subsection{The NO$_2^{-}$:NO$_2^{-}$ intermolecular potentials}

Similarly, we show in Fig.~\ref{fig:no2-no2}
the NO$_2^{-}$-NO$_2^{-}$ intermolecular potentials using the same
four methods. The configurations are chosen as follows: the two
NO$_2^{-}$ molecules are initially parallel at their experimental
low-temperature separation ($1.82$ \AA , $2.83$ \AA , $2.69$ \AA )
and then one of them rotates around each of the $x$, $y$, $z$ axes
through its center of mass, as shown in Figs.~\ref{fig:no2-no2}(a)
through \ref{fig:no2-no2}(c). The results of models I-III agree
reasonably well with the \textit{ab initio} calculations. On
closer examination, Models II and III fit better to the \textit{ab
initio} results than I. We believe this is due to the fact that
the two nitrite ions are separated so far that the main
contribution to the intermolecular potential is mostly
electrostatic. Based on the experience acquired in Section IIB,
FPA usually gives a better description of the rigid NO$_2^{-}$
molecule than MPA, hence the results obtained from models II and
III which employ the FPA are considerably better. In addition, the
short range interaction is also small, thus the difference between
model II and III is rather small.

Summarizing this section, in spite of the presence of electronic
polarization when two molecules are brought closer, the
intermolecular potentials for the NO$_2^{-}$:Na$^{+}$ and
NO$_2^{-}$:NO$_2^{-}$ dimers could be correctly reproduced using
models I and II. Furthermore, model II appears to give better
description of NO$_2^-$ than model I as shown in
Fig.~\ref{fig:no2-no2}, particularly Figs.~\ref{fig:no2-no2}(a)
and \ref{fig:no2-no2}(c).

%The resultant short-range pair potentials are summarized in
%Fig.~\ref{fig:pair}.

\section{Molecular Dynamics Simulations}

After the potential energy surface for NaNO$_2$ has been obtained,
we are prepared to undertake MD simulations. Long-range Coulombic
interaction in the crystal is represented by electrostatic
interaction among point charges calculated from the population
analysis, while each of the short-range pair potentials are fitted
by using an exponential-polynomial function accurate within 0.1\%.
\cite{hardy:kno3}  In the following description,
the $x$, $y$, $z$ directions correspond to the crystallographic $a$, $%
b$, $c$ directions of NaNO$_2$, respectively, see
Fig.~\ref{fig:nano2}.

\subsection{Lattice relaxation}

\begin{table}%[H] add [H] placement to break table across pages
\caption{\label{table:par}%
Experimental and theoretical structural parameters for the Im2m
phase (III) of NaNO$_2$. Lattice constants are given in \AA.}
\begin{ruledtabular}
\begin{tabular}{lccccc}
Parameters & Experiment\footnotemark[1] & Model I & Model II & Model III \\
\hline
a               &  3.5180 &   3.3889  &   3.5013  &   3.7353 \\
b               &  5.5350 &   5.4542  &   5.5485  &   5.4257 \\
c               &  5.3820 &   4.9254  &   4.8403  &   4.9669 \\
y/b of N (2a)   &  0.5437 &   0.0498  &   0.0433  &   0.0586 \\
y/b of Na (2a)  &  0.0781 &   0.5537  &   0.5492  &   0.5437 \\
y/b of O (4d)   & -0.0443 &  -0.0704  &  -0.0740  &  -0.0610 \\
z/c of O (4d)   &  0.1962 &   0.2111  &   0.2151  &   0.2098 \\
\end{tabular}
\end{ruledtabular}
\footnotetext[1]{From X-ray diffraction experiments at 120 K,
see Ref.~\onlinecite{nano2:okuda}.}%
\end{table}

Before we proceed with the molecular dynamics simulations, we
perform lattice relaxation for the ferroelectric structure of
NaNO$_2$ both with and without the $Im2m$ space group symmetry
constraints. This relaxation procedure provides the crystal
structure with zero force on each atom, that is an energy
extremum; it also produces a test to the PES because the resultant
structures have to agree reasonably with the experimental data for
further simulations to be reliable. We perform both static and
dynamic relaxations: the static one is an application of the
Newton-Raphson algorithm and usually results in finding a local
minimum of the energy, and the dynamic one is a simulated
annealing calculation for overcoming that limitation. We start the
static lattice relaxation with the experimental parameters. In
Table~\ref{table:par} we present the lattice and basis parameters
deduced from the experiments and static relaxation. In all cases,
the static relaxation produced essentially the same structure that
strongly resembles the experimental structure. Most of the lattice
constants in the relaxed structure are shorter than the
experimental values (by 3.7\%, 1.5\%, and 8.5\% for $a$, $b$, and
$c$, respectively, in model I, and by 0.5\%, -2.4\%, and 10\% for
$a$, $b$, and $c$, respectively, in model II). Hence the
calculated volume is smaller than the experimental one by 13\% for
model I and 10\% for model II, a common feature for simulations
using the GK model, which will be addressed in more detail in the
next subsection.

Next, we go on to relax the statically relaxed crystal structure
to zero temperature using a simulated annealing algorithm, in
which the amount of kinetic energy in the molecules slowly
decreases over the course of the simulation. We find that the
(zero temperature) ground states in models I and II are close to
the statically relaxed structures, whereas there are substantial
changes taking place in model III. By monitoring the orientations
of the nitrite ions, we find that the ground structure in model
III, still orthorhombic with $a=3.90494$ \AA, $b=4.8441$ \AA, and
$c=5,0770$ \AA, is ferroelectric with the dipole moments of
NO$_2^{-}$ aligned along the $a$ axis rather than the experimental
$b$ axis. So we conclude that the PES given by models III did not
reflect reality. This concurs with the previous discussion
(Section III) on the intermolecular potentials. In the following
we use only models I and II to simulate the phase transition in
NaNO$_2$.

\subsection{MD simulations}

\begin{figure}[t]
\includegraphics*{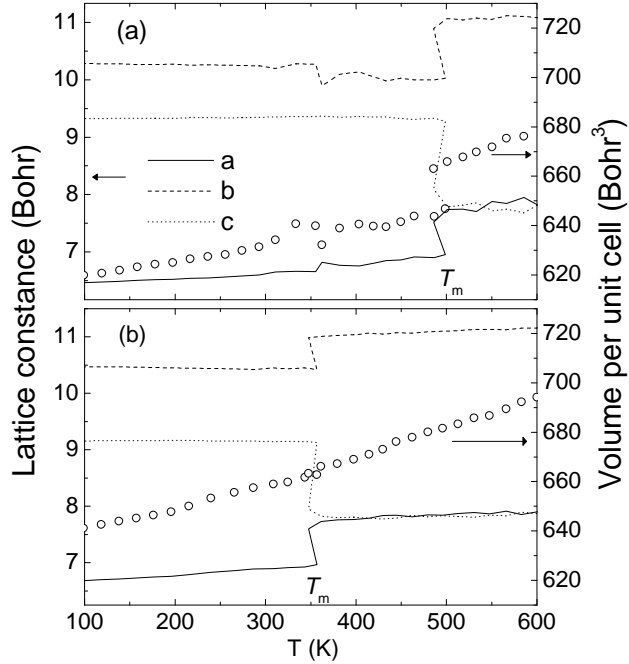}%
\caption{\label{fig:abc}%
Temperature variation of lattice constants $a$, $b$, $c$ (solid,
dashed, dotted lines, respectively; left scale) and volume of the
unit cell (open circles; right scale) for (a) model I and (b)
model II. }
\end{figure}

\begin{figure}
\includegraphics*{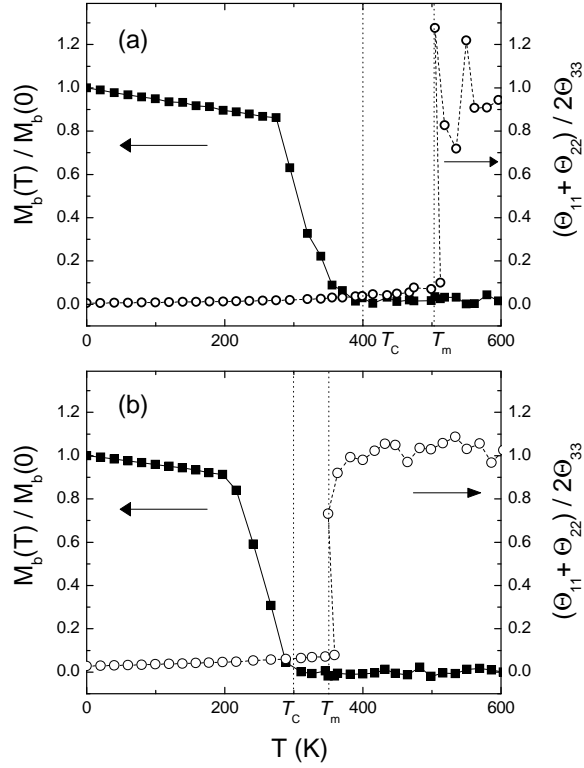}%
\caption{\label{fig:fe}%
Mean dipole moment $M_b(T)$ and quadrupole moment $\Theta$ of the whole NaNO$%
_2$ crystal as a function of temperature for the MD runs for (a)
model I and (b) model II. }
\end{figure}

\begin{figure}
\includegraphics*{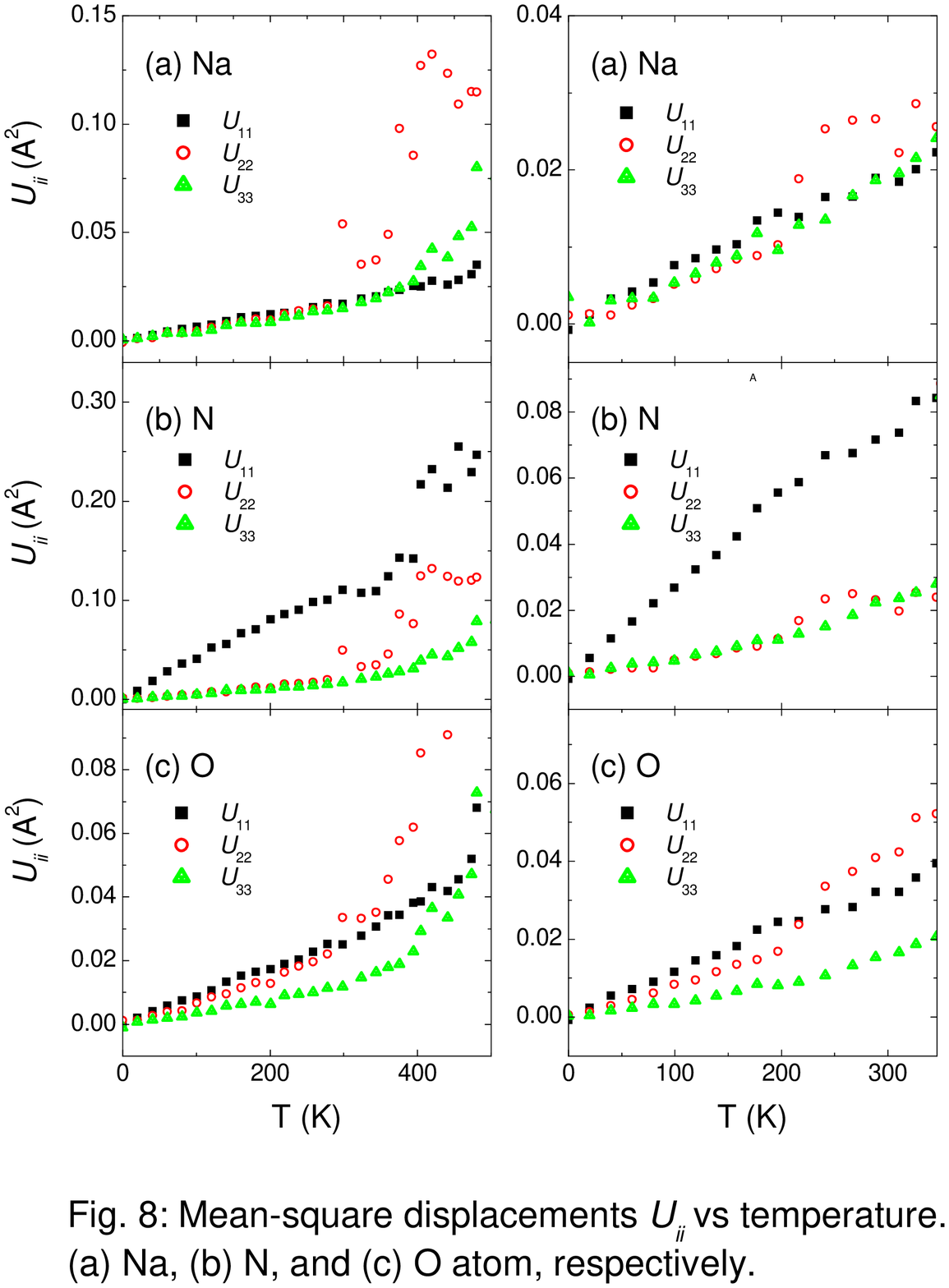}%
\caption{\label{fig:uu}%
Diagonal elements of the mean-square atomic displacements $U_{ii}$
vs temperature. (a) Na, (b) N, and (c) O atom. }
\end{figure}

\begin{figure}[t]
\includegraphics*{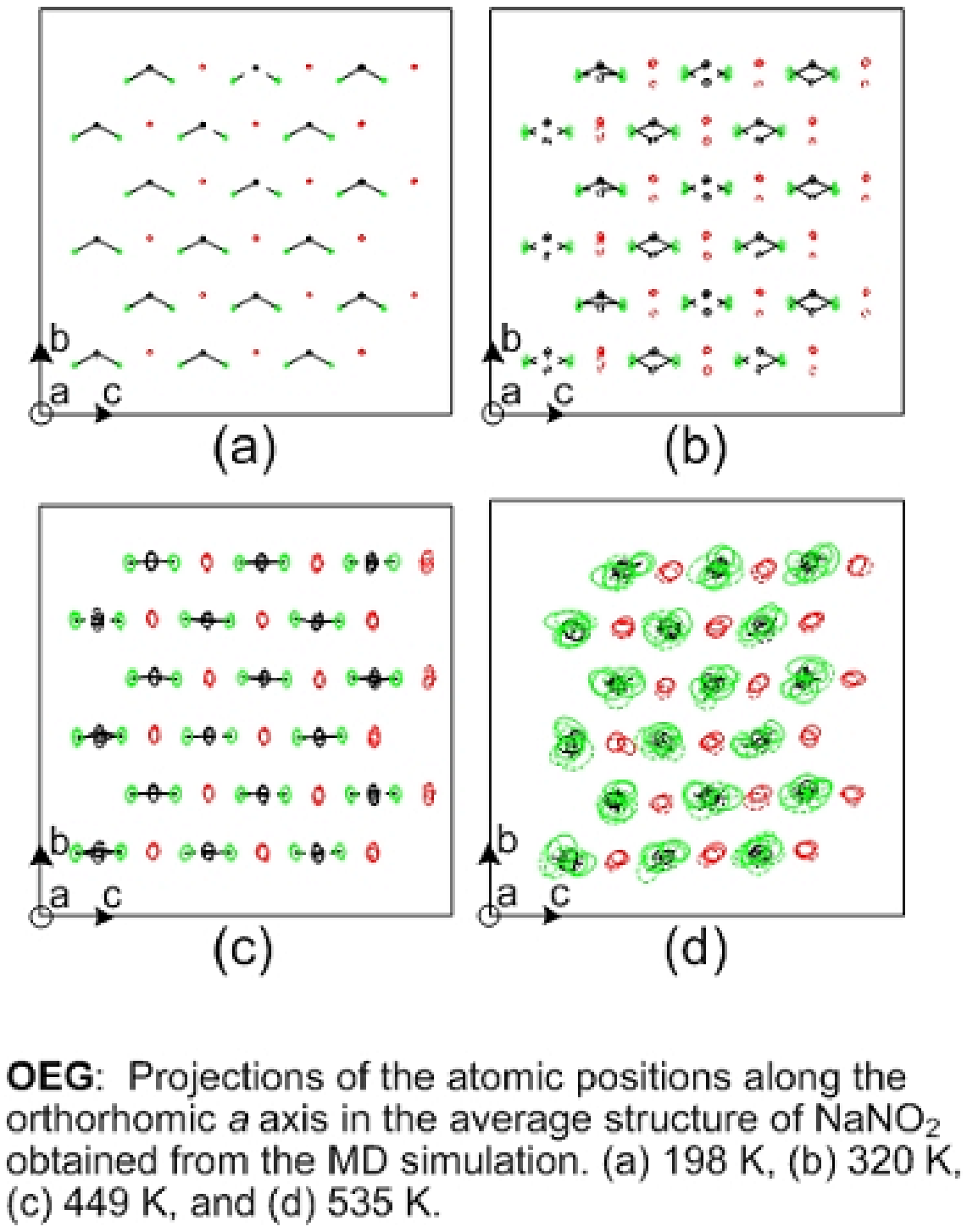}%
\caption{\label{fig:ellipsoid}%
Atomic positions of NaNO$_2$ viewed from the $a$ direction
obtained
from the MD simulation for model I at (a) $T=198$ K, (b) $T=320$ K, (c) $%
T=449$ K, and (d) $T=535$ K. }
\end{figure}

Using the isoenthalpic, isobaric ensemble, our MD simulation is
started with a zero-temperature zero-pressure orthorhombic cell
($4a\times 4b\times 4c$), consisting of 512 atoms. Periodic
boundary conditions are imposed to simulate an infinite crystal.
The MD calculations are carried out in the Parrinello-Rahman
scheme \cite{parrinello-rahman:80} which allows both the volume
and the shape of the MD cell to vary with time. The calculation of
the energies and forces was handled by the Ewald method. A time
step of 0.002 ps was used to integrate the equations of motion. In
our heating runs, we raise the temperature of the sample in
stages, 20 K each time, up to 1000 K. At each stage, the first
2000 time steps were employed to equilibrate the system, then
10000 time steps were collected for subsequent statistical
analysis. Since our simulations employ periodic boundary
conditions, we cannot distinguish the incommensurate structure
(i.e., phase II of solid NaNO$_2$).

\ignore{In Fig.~\ref{fig:abc} we present the temperature variation
of lattice constants and
volume of the unit cell. %Two noticeable changes in volume take
%place around $T_{\mathrm{C}}$ and $T_{\mathrm{m}}$;
The crystal structure remains orthorhombic around $T_{\mathrm{C}}$
and changes to tetragonal around $T_{\mathrm{m}}$. }

In Figs.~\ref{fig:abc} through \ref{fig:ellipsoid}, we demonstrate
that as the MD cell is heated, it undergoes two phase transitions:
in the first one, the system retains its orthorhombic structure
with a change of space group from $Im2m$ to $Immm$, in agreement
with the experiments. The critical temperature $T_{\mathrm{C}}$ is
around $400$ K for model I and $313$ K for model II, respectively.
In the second transition, the crystal structure violently changes
from orthorhombic to tetragonal at temperature ($T_{\mathrm{m}}$)
which is around $500$ K for model I and $350$ K for model II, as
shown in Fig.~\ref{fig:abc}. However, we argue that the crystal
has already melted before this type of transition could be
observed in reality.
%In other words, the second transition signals the onset of melting. \cite
%{hardy:nano2}

To investigate the mechanism of the polarization reversal of NO$%
_2^{-}$, we monitor the crystal polarization and display the
results in Fig.~\ref{fig:fe}. Let the dipole moment of anion $i$
be $\mathbf{m}_i$ and the quadrupole
moment be $\mathbf{\vartheta}_i$ calculated by using the point charges on the $%
N$ and $O$ atoms. Then the mean dipole moment per anion at
temperature $T$ is $\mathbf{M(}T\mathbf{)}=\sum_i\left\langle \mathbf{m}%
_i\right\rangle /N$ where $N=128$ is the number of NO$_2^{-}$ in
the MD cell and the brackets denote an average over the MD run. In
addition, we define the antiferroelectric polarization as
$\mathbf{Q}=\sum_i\exp (-\mathbf{\pi
\cdot R}_i)\left\langle \mathbf{m}_i\right\rangle /N$ where $\mathbf{\pi }%
=(\pi ,\pi ,\pi )$ and $\mathbf{R}_i$ is the lattice vectors
associated with the $i$-th ion. Within our statistical uncertainty
we find over all temperature range $M_x=M_z=\mathbf{Q}=0$, while
$M_y(T<T_{\mathrm{C}})>0$ and $M_y(T>T_{\mathrm{C}})=0$. This fact
confirms that the transition taking place at $T_{\mathrm{C}}$ is
the ferroelectric-paraelectric phase transition. Furthermore, we
calculated the mean quadrupole moment $\mathbf{\Theta
}=\sum_i\left\langle \mathbf{\vartheta }_i\right\rangle /N$. When
the dipole vector of a NO$_2^{-}$ is aligned along the $b$ axis,
$\vartheta _{xx}\simeq 0.00$, $\vartheta _{yy}\simeq -0.04$,
$\vartheta _{zz}\simeq -4.49$ for model I and $\vartheta
_{xx}\simeq 0.00$, $\vartheta _{yy}\simeq -0.20$, $\vartheta
_{zz}\simeq -3.52$ for model II; thus $(\Theta _{xx}+\Theta
_{yy})/2\Theta _{zz}\ll 1$. This relation holds as the NO$_2^{-}$
ion rotates around the $c$ axis; nevertheless, one would expect
$(\Theta _{xx}+\Theta _{yy})/2\Theta _{zz}=1$ when the NO$_2^{-}$
ion rotates without directional preference. The
fact that $(\Theta _{xx}+\Theta _{yy})/2\Theta _{zz}\ll 1$ for $T<T_{\mathrm{%
m}}$ (Fig.~\ref{fig:fe}) reveals that the NO$_2^{-}$ anions rotate
primarily about the $c$ axis. When $T>T_{\mathrm{m}}$, $(\Theta
_{xx}+\Theta _{yy})/2\Theta _{zz}\simeq 1$, i.e., NaNO$_2$ becomes
an orientational liquid.

Further, in Fig.~\ref{fig:uu} we show the mean-square atomic
displacements $U_{ii}=\left\langle
u_i^2\right\rangle $ where $i=1,2,3$ denotes the displacements along the $%
a,b,c$ axes, respectively. Different models of NO$_2^{-}$ reversal
are expected to satisfy the following conditions: (1) rotation
around the $c$
axis: $U_{22}(\mathrm{N}),U_{33}(\mathrm{N})<U_{11}(\mathrm{N})$ and $U_{22}(%
\mathrm{O}),U_{33}(\mathrm{O})<U_{11}(\mathrm{O})$; (2) rotation around the $%
a$ axis: $U_{11}(\mathrm{N}),U_{22}(\mathrm{N})<U_{33}(\mathrm{N})$ and $%
U_{11}(\mathrm{O}),U_{33}(\mathrm{O})<U_{22}(\mathrm{O})$.  This
figure relates to recent X-ray experiments which used the same
quantities to investigate the polarization reversal mechanism.
\cite {nano2:gohda,nano2:komatsu}  The experiments confirmed that
the first condition holds for both ferroelectric and paraelectric
phases. Another important feature revealed by the experiments is
that $U_{22}(\mathrm{Na}),U_{33}(\mathrm{Na})<U_{11}(\mathrm{Na})$
in the
ferroelectric phase, whereas $U_{11}(\mathrm{Na}),U_{33}(\mathrm{Na})<U_{22}(%
\mathrm{Na})$ in the paraelectric phase. That is, $U_{11}(\mathrm{Na})$ and $%
U_{22}(\mathrm{Na})$ are reversed across $T_{\mathrm{C}}$. These features
are reproduced in Fig.~\ref{fig:uu} with exception of $U_{11}(\mathrm{O}),U_{33}(%
\mathrm{O})<U_{22}(\mathrm{O})$ in the paraelectric phase. This means the NO$%
_2^{-}$ motions in our simulations are more mobile than those in
the real crystal, rendering the simulated transition temperatures
lower than the experimental values of $T_{\mathrm{C}}\simeq 437$ K
and the melting temperature $550$ K. In other words, the barriers
to NO$_2^{-}$ rotation in our models are too small.

In addition, in Fig.~\ref{fig:ellipsoid}, we show the average
crystal structures of NaNO$_2$ at different temperatures. The
ellipsoids in these pictures represent the root-mean-square
deviations of the atoms from their average positions and thus
indicate the thermal motions of these atoms. The $c$-axis rotation
mode can be clearly seen, particularly in
Fig.~\ref{fig:ellipsoid}(c).

According to the calculations described in the previous section,
model II generally gives a better description of free NO$_2^{-}$
and the intermolecular potential energies than model I. However,
the simulation based on model I matches closer the experimental
results than that based on model II, that is, $T_{\mathrm{C}%
} $ and $T_{\mathrm{m}}$ predicted by model I are closer to
experiment, this indicates that the crystal fields and
polarization effects are particularly important for quantitatively
studying the NaNO$_2$ system, where the ferroelectric structure is
considerably stabilized by anion polarization effects. Actually,
the MPA employed by model I does not preserve the \emph{ab initio}
dipole moment of free NO$_2^{-}$. Rather, it overestimates the
dipole moment by 120\%, thus leading to higher NO$_2^-$ rotational
barriers than those predicted by model II, which in turn raises
the transition temperature and provides a better simulation in
comparison with the experiments.

It is worth mentioning the less desirable agreement between
theoretical and experimental volumes, namely, the 13\% discrepancy
for model I and 10\% for model II. To address this we make one
simple change: by following Waldman and
Gordon,\cite{gordon:waldman:scale} we increase the kinetic energy
term in the Gordon-Kim potentials by 5\%, this reduces the
discrepancy to 9\% for model I and 6\% for model II. Having done
this we rerun the MD to obtain values of $T_c$ of 360 K for model
I and 303 K for model II. While this change worsens the value for
model I, the value for model II is virtually unchanged. And in
both cases the transition mechanism is unaltered. Thus the slight
hardening of the short-range potentials removes most of the volume
discrepancies. However, there is no material change in the
mechanism of the phase transition. This robustness of the results
with respect to minor variations in the potential demonstrates
that our basic conclusion remain valid.

\subsection{Rotational barriers}

\begin{figure}
\includegraphics*{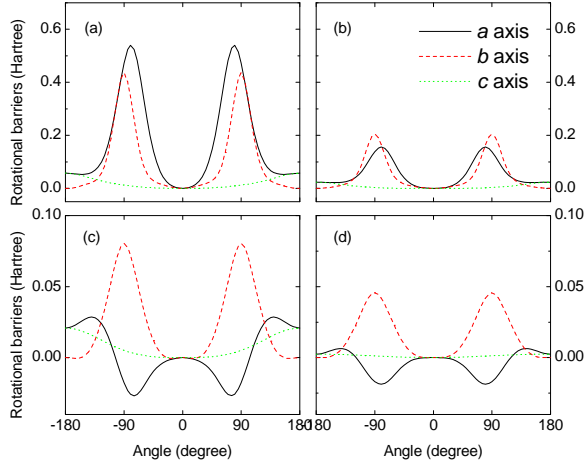}%
\caption{\label{fig:barrier}%
Rotational barriers of one of the two nitrite ions in the unit
cell around the $a$, $b$, and $c$ axes with its center of mass
fixed. (a)(b) for the GK model, (c)(d) for the point charge model.
Left and right panels are for model I and II, respectively. }
\end{figure}

Based on the previous simulation results, the order-disorder phase
transition in NaNO$_2$ involves the rotation of the nitrite ions.
We devise a scheme to calculate the three rotational barriers for
NO$_2^{-}$ around the crystallographic $a$, $b$, and $c$ axes with
its center of mass fixed: we start from the experimental
ferroelectric structure \cite{nano2:kay} as the ground state and
calculate its energy difference with one of the two nitrite ions
in the unit cell being rotated about the respective axis. The
results are shown in Figs.~\ref{fig:barrier}(a) and
\ref{fig:barrier}(b).
The bottom of each barrier, zero angle, is in the ferroelectric
structure. For both models I and II, the rotation around the
$c$-axis has an energy barrier 5-10 times smaller than those of
the other rotations, which is a characteristic of nitrites.
\cite{hardy:nitrite} Hence, our calculations unambiguously reveal
that the reorientation of NO$_2^{-}$ in the paraelectric phase
occur essentially by rotations around the $c$ axis. In addition,
the barriers calculated in model I are higher than that in model
II, confirming that the transition temperature in model I will be
higher than in model II.

In Figs.~\ref{fig:barrier}(c) and \ref{fig:barrier}(d), we also
plot the contribution to the rotational barriers purely from the
electrostatic interaction, that is, in the point-charge model with
the short range forces deleted. Comparing
Fig.~\ref{fig:barrier}(a) with Fig.~\ref{fig:barrier}(c), or
Fig.~\ref{fig:barrier}(b) with Fig.~\ref{fig:barrier}(d), we
notice that the point-charge model gives rise to a quite different
rotation barrier about the $a$ axis: it bottoms at about $\pm
90^{\circ }$ and is lower than the ground state energy due to the
omission of short-range interactions, which comes from overlap of
the charge cloud of an atom with those of its neighbors.

\section{Crystal-field effects}

In this section, we investigate the crystal-field effects on the
NO$_2^{-}$ ion, which include electrostatic interaction from the
background lattice, overlap compression of the NO$_2^{-}$ charge
cloud through interaction with its neighbors, and charge-transfer
between molecules which is usually small in ionic crystals. In the
studies of monatomic ions \cite {alkali_halide:fowler} and
cyanides \cite{nacn:fowler}, Fowler \textit{et al.} showed that
these effects could be successfully described by embedding the ion
of interest, or a cluster consisting of the ion and its first
shell of neighbors, into a lattice of point charges.

%\subsection{Polarizable ion models}

We therefore perform HF calculations based on the following two
models. \cite {alkali_halide:fowler}  First, the crystal field of
ferroelectric NaNO$_2$ is simulated by placing the nitrite ion at
the center of a $4\times 4\times 4$ orthorhombic point charge
lattice with spacings equal to the experimental lattice
parameters. Charges in the faces of the lattice are scaled to
maintain overall neutrality. All anions except the central
NO$_2^{-}$ are represented by single point charges. Hence, there
are 174 point charges surrounding the NO$_2^{-}$ ion. Calculations
of this type are referred to as CRYST. Obviously, in the CRYST
calculation we take into account only the crystal-field effect
arising purely from electrostatic interaction.

At the next level of sophistication, we replace the six nearest
positive charges of the central NO$_2^{-}$ in the above lattice by
the Na cations. Calculations of this type are referred to as
CLUST. In both CRYST and CLUST calculations, the geometrical
structure of the NO$_2^-$ is fixed at its experimental values. We
employ the same basis set, D95*, for the in-crystal NO$_2^-$ ion
as for the free NO$_2^{-}$ ion. In order to keep the CLUST
calculations to a manageable size, we use the minimal basis set,
STO-3G, for the Na$^+$ ions unless specified. The cations,
however, are relatively insensitive to the crystal environment and
they are included here only to account for their compressing
effect on the NO$_2^{-}$ wave functions. We find that adding extra
basis functions to Na$^+$ will not change the results
significantly.

%\subsection{Dipole moment of in-crystal NO$_2^{-}$}

\begin{figure}
\includegraphics*{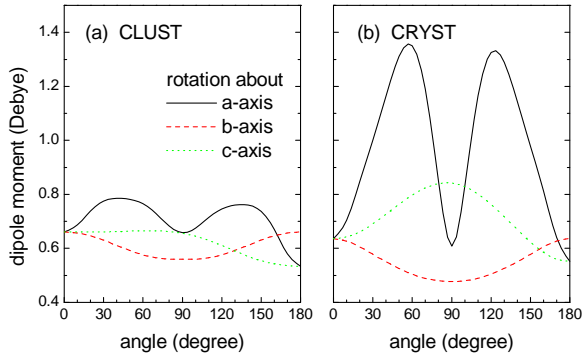}%
\caption{\label{fig:polarizable}%
Dipole moment of the central NO$_2^{-}$ in a $4\times 4\times 4$
lattice as it rotates around the $a,b,c$ axes through the center
of its (Na$^+$)$_6$ cage. (a) CLUST and (b) CRYST. }
\end{figure}

The in-crystal NO$_2^{-}$ initially points in the $b$ direction as
in the ferroelectric phase of NaNO$_2$ \cite{nano2:kay}. Its
dipole moment is 0.636 (CRYST) and 0.661 (CLUST) Debye, close to
that in the free ion model, 0.661 Debye. Thus it appears that the
crystal-field effect is small for this configuration.
Subsequently, we rotate the NO$_2^{-}$ about the $a$, $b$, and $c$
axes and calculate the dipole moment of the rotated NO$_2^{-}$.
The rotation center is taken to be the center of a (Na$^+$)$_6$
cage formed by the 6 neighboring sodium ions of the central
NO$^-_2$,\cite{nano2:michel} (0,0.279 \AA, 0) in the coordinate
convention of Fig.~\ref{fig:nano2}.  The CLUST and CRYST results
are depicted in Figs.~\ref{fig:polarizable}(a) and
\ref{fig:polarizable}(b), respectively.
Clearly, the dipole moment of the NO$_2^{-}$ changes considerably
as it rotates, indicating strong crystal field effects on the
reorientation of the NO$_2^{-}$. Since the electron density of the
NO$_2^{-}$ is compressed by its 6 neighboring sodium cations, the
variation in magnitude of its dipole moment is smaller in the
CLUST model than that in the CRYST model. The electron cloud of
the NO$_2^{-}$ is most and least variable when it rotates around
the $a$ and $b$ axes, respectively; for the $c$-axis rotation,
which has the lowest rotational barrier, the dipole moment goes
down from 0.661 Debye at $0^\circ$ to 0.534 Debye at $180^{\circ
}$ in the CLUST model, as shown Fig.~\ref{fig:polarizable}(a).

In the context of population analysis, increase of the dipole
moment of NO$^-_2$ implies that more electrons are distributed on
the O atom, i.e., electrons are flowing from the nitrogen atom to
the oxygen atoms. Conversely, decrease of the dipole moment
indicates a reversal in electron transfer. Therefore, we have
demonstrated considerable \emph{intramolecular} charge-transfer,
although the intermolecular charge-transfer is usually small in
ionic crystals. This intramolecular charge-transfer could be
further elucidated based on the language of molecular orbitals
(MOs): in the minimal basis set of atomic orbitals (AOs) on
nitrogen and oxygen, each atom of the NO$_2^-$ molecule
contributes one $p$-orbital perpendicular to the molecular plane.
Thus their linear combinations, which are determined by the
crystal field, form three different $\pi$ MOs extended over the
entire molecule, thus leading to the above-mentioned
intramolecular charge-transfer. Note that the mean HF
polarizability of free NO$_2^{-}$, $14.156$ in the atomic units,
is much higher than that of free Na$^+$, $0.343$, calculated by
using the 6-31* basis. Therefore, it is reasonable to expect that
NO$_2^{-}$ in solid NaNO$_2$ encountering different crystal-field
environments as it rotates, redistributes its charge among the
three constituent atoms to lower its energy.

\begin{figure}
\includegraphics*{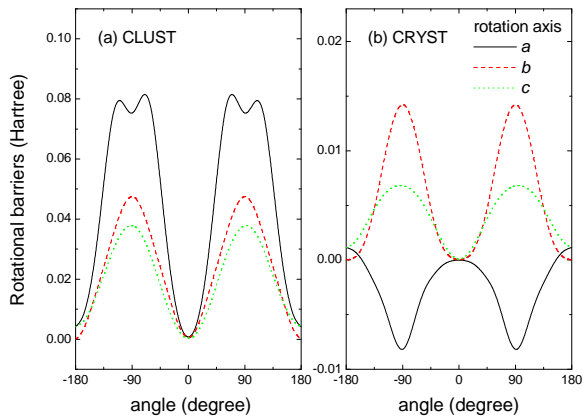}%
\caption{\label{fig:barrier444}%
{\em ab initio} barriers to rotation of the central NO$_2^{-}$ in
a $4\times 4\times 4$ lattice around the $a$, $b$, $c$ axes
through the center of its (Na$^+$)$_6$ cage. (a) CLUST and (b)
CRYST. }
\end{figure}

In Fig.~\ref{fig:barrier444}
we show that the rotational barriers of the central NO$_2^{-}$. On
the whole, The CLUST results are similar to those in
Figs.~\ref{fig:barrier}(a) and \ref{fig:barrier}(b), while the
CRYST results are similar to those in Figs.~\ref{fig:barrier}(c)
and \ref{fig:barrier}(d). The reason is that in the CRYST
calculations we consider only the crystal-field effects
originating purely from electrostatic interactions with the
background point charges, similar to the point charge model used
to obtain Figs.~\ref{fig:barrier}(c) and \ref{fig:barrier}(d).
Whereas, in the CLUST calculations, overlap compression is also
taken into account, which is reflected in
Figs.~\ref{fig:barrier}(a) and \ref{fig:barrier}(b) as inclusion
of short range repulsion. In Fig.~\ref{fig:barrier444}(a), the
barrier to the $c$-axis rotation is the lowest, but comparable
with that to the $b$-axis rotation. This feature is caused by the
fact that all background anions are represented by single point
charges in our CLUST and CRYST calculations; however, we
anticipate that restoring multipole moments of these background
anions would increase the barrier difference among rotations about
the $a$, $b$, and $c$ axes.

\begin{figure}
\includegraphics*{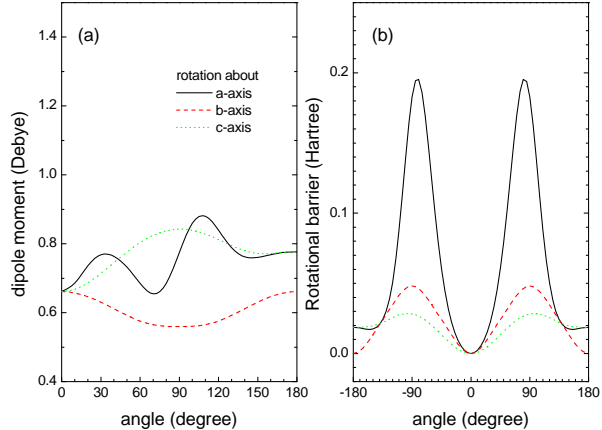}%
\caption{\label{fig:crystal}%
In the CLUST model, NO$_2^{-}$  rotates around the $a,b,c$ axes
through its center of mass. (a) dipole moment of NO$_2^{-}$, and
(b) rotational barriers. }
\end{figure}

To further elaborate the electronic polarization effect on
NO$^-_2$ arising from the crystal environment, we change the
rotation center to the center of mass of the NO$^-_2$ which was
assumed in the model by Kremer and Siems. \cite{nano2:kremer} In
this case, the dipole moment of NO$^-_2$ and the rotational
barriers are presented in Figs.~\ref{fig:crystal}(a) and
\ref{fig:crystal}(b), respectively.
The dipole moment changes in a different way from that shown in
Fig.~\ref{fig:polarizable}(a). In particular, the dipole moment in
the ferroelectric phase ($0^\circ$-rotation) is larger than at
$180^\circ$-rotation in Fig.~\ref{fig:polarizable}(a), while it is
smaller in Fig.~\ref{fig:crystal}(a). On the other hand, there is
no qualitative discrepancy between Fig.~\ref{fig:crystal}(b) and
Fig.~\ref{fig:barrier444}(a); the main differences are: the
rotational barrier about the $a$ axis has risen by 135\% while the
barrier about the $c$ axis is depressed by 26\%. This means that
the order of rotational barriers is enhanced by the change of the
rotation center.

Although strong crystal field effects have been revealed by these
\emph{ab initio} calculations, the rotational barriers obtained
from the polarizable-ion models are in qualitative agreement with
those from the rigid-ion models, confirming that the rigid-ion
model is capable of describing the phase behavior in NaNO$_2$.

%To finish the present investigation, we perform the MD simulation
%using the rigid-ion model with the CRYST NO$_2^{-}$ at $0^\circ$
%(ferroelectric phase) being the rigid ion. In such a treatment the
%crystal-field effect is partially taken into account. The FPA and
%pairwise additive approximation are used. All results are similar
%to the previous calculations using the free ion approximation, as
%shown in Table~\ref {table:par} and Fig.~\ref{fig:fe444}.
%\begin{figure}
%\includegraphics*{fig12.eps}%
%\caption{\label{fig:fe444}%
%Mean dipole moment $M_b(T)$ and quadrupole moment $\Theta$ of the
%whole NaNO$_2$ crystal as a function of temperature for the MD runs
%for the in-crystal NO$_2^{-}$ (CRYST) model. }
%\end{figure}
%The resulting critical temperatures $T_{\mathrm{C}}$ $\sim 330$ K
%and $T_{\mathrm{m}}\sim 380$ K are better than those in model II
%which also employed the FPA, but still worse than in model I. We
%already shew that in model I the dipole moment of NO$_2^{-}$ is
%overestimated. It is this overestimation that stabilizes the
%ferroelectric structures of NaNO$_2$ within a rigid-ion model,
%which simulates the anion polarization effects on stabilization of
%the crystal structures.
%In retrospect, in our polarizable anions models, the ferroelectric
%configuration of $0^\circ$ rotation of NO$^-_2$ has a larger dipole
%moment than the $180^\circ$-rotation phase.

\section{Concluding remarks}

We have presented MD simulations of NaNO$_2$ using a hybrid
\emph{a priori} method consisting of \emph{ab initio} calculations
and Gordon-Kim electron gas theory to analytically calculate the
crystal potential surface. This method has been carefully examined
by using different population analysis methods. We have carried
out \emph{ab initio} Hartree-Fock calculations of the
intermolecular interactions for NO$_2^{-}$:Na$^+_{}$ and
NO$_2^{-}$:NO$_2^{-}$ dimers and concluded that the pair
potentials of the rigid-ion model can correctly reproduce the
\emph{ab initio} results. We demonstrated that a rigid-ion model
is capable of describing phase behavior in solid NaNO$_2$.

The crystal-field effects on the NO$_2^{-}$ ion are also addressed
in two polarizable-ion models for which the ferroelectric phase of
NaNO$_2$ was found to have a larger dipole moment of NO$_2^{-}$
than the $180^\circ$-rotation phase. Remarkable internal
charge-transfer effect is found to be stabilizing the crystal
structure of NaNO$_2$.
%
%
%In our MD simulations, three rigid-ion models have been studied:
%free ions with the MPA, free ions with the FPA,
%and in-crystal ions at the ferroelectric configuration
%with the FPA. Among them, the first one that overestimates the
%dipole moment of NO$_2^{-}$ gives rise to the most
%
%
In our MD simulations, two rigid-ion models using MPA and FPA,
respectively, have been studied.  The one using MPA, which
overestimates the dipole moment of NO$_2^{-}$, gives rise to the
more comparable results with the experiments, since such
overestimation also stabilizes the crystal structure, thus mimics
the anion polarization effect. To quantitatively simulate
NaNO$_2$, a more elaborate polarizable-ion model is needed.

%\appendix
%\section{}

\begin{acknowledgments}
Helpful discussions with Dr. L. L. Boyer are gratefully
acknowledged. This work was supposed by the Nebraska Research
Initiative, the Nebraska EPSCoR-NSF Grant EPS-9720643, and
Department of the Army Grants DAAG 55-98-1-0273 and DAAG
55-99-1-0106. W. N. M. is grateful for the support from the Office
of Naval Research.
\end{acknowledgments}


\begin{thebibliography}{41}
\expandafter\ifx\csname natexlab\endcsname\relax\def\natexlab#1{#1}\fi
\expandafter\ifx\csname bibnamefont\endcsname\relax
  \def\bibnamefont#1{#1}\fi
\expandafter\ifx\csname bibfnamefont\endcsname\relax
  \def\bibfnamefont#1{#1}\fi
\expandafter\ifx\csname citenamefont\endcsname\relax
  \def\citenamefont#1{#1}\fi
\expandafter\ifx\csname url\endcsname\relax
  \def\url#1{\texttt{#1}}\fi
\expandafter\ifx\csname urlprefix\endcsname\relax\def\urlprefix{URL }\fi
\providecommand{\bibinfo}[2]{#2}
\providecommand{\eprint}[2][]{\url{#2}}

\bibitem[{\citenamefont{Sawada et~al.}(1958)\citenamefont{Sawada, Nomura,
  Fujii, and Yoshida}}]{nano2:sawada}
\bibinfo{author}{\bibfnamefont{S.}~\bibnamefont{Sawada}},
  \bibinfo{author}{\bibfnamefont{S.}~\bibnamefont{Nomura}},
  \bibinfo{author}{\bibfnamefont{S.}~\bibnamefont{Fujii}}, \bibnamefont{and}
  \bibinfo{author}{\bibfnamefont{I.}~\bibnamefont{Yoshida}},
  \bibinfo{journal}{Phys.\ Rev.\ Lett.} \textbf{\bibinfo{volume}{1}},
  \bibinfo{pages}{320} (\bibinfo{year}{1958}).

\bibitem[{\citenamefont{Lines and Glass}(1977)}]{lines:glass}
\bibinfo{author}{\bibfnamefont{M.~E.} \bibnamefont{Lines}} \bibnamefont{and}
  \bibinfo{author}{\bibfnamefont{A.~M.} \bibnamefont{Glass}},
  \emph{\bibinfo{title}{Principles and Applications of Ferroelectrics and
  Related Materials}} (\bibinfo{publisher}{Clarendon},
  \bibinfo{address}{Oxford}, \bibinfo{year}{1977}).

\bibitem[{\citenamefont{Fokin et~al.}(2002)\citenamefont{Fokin, Kumzerov,
  Okuneva, Naberezhnov, Vakhrushev, Golosovsky, and Kurbakov}}]{nano2:fokin}
\bibinfo{author}{\bibfnamefont{A.~V.} \bibnamefont{Fokin}},
  \bibinfo{author}{\bibfnamefont{Y.~A.} \bibnamefont{Kumzerov}},
  \bibinfo{author}{\bibfnamefont{N.~M.} \bibnamefont{Okuneva}},
  \bibinfo{author}{\bibfnamefont{A.~A.} \bibnamefont{Naberezhnov}},
  \bibinfo{author}{\bibfnamefont{S.~B.} \bibnamefont{Vakhrushev}},
  \bibinfo{author}{\bibfnamefont{I.~V.} \bibnamefont{Golosovsky}},
  \bibnamefont{and} \bibinfo{author}{\bibfnamefont{A.~I.}
  \bibnamefont{Kurbakov}}, \bibinfo{journal}{Phys.\ Rev.\ Lett.}
  \textbf{\bibinfo{volume}{89}}, \bibinfo{pages}{175503}
  (\bibinfo{year}{2002}).

\bibitem[{\citenamefont{Gohda et~al.}(2000)\citenamefont{Gohda, Ichikawa,
  Gustafsson, and Olovsson}}]{nano2:gohda}
\bibinfo{author}{\bibfnamefont{T.}~\bibnamefont{Gohda}},
  \bibinfo{author}{\bibfnamefont{M.}~\bibnamefont{Ichikawa}},
  \bibinfo{author}{\bibfnamefont{T.}~\bibnamefont{Gustafsson}},
  \bibnamefont{and} \bibinfo{author}{\bibfnamefont{I.}~\bibnamefont{Olovsson}},
  \bibinfo{journal}{Phys.\ Rev.\ B} \textbf{\bibinfo{volume}{63}},
  \bibinfo{pages}{014101} (\bibinfo{year}{2000}).

\bibitem[{\citenamefont{Komatsu et~al.}(1988)\citenamefont{Komatsu, Itoh, and
  Nakamura}}]{nano2:komatsu}
\bibinfo{author}{\bibfnamefont{K.}~\bibnamefont{Komatsu}},
  \bibinfo{author}{\bibfnamefont{K.}~\bibnamefont{Itoh}}, \bibnamefont{and}
  \bibinfo{author}{\bibfnamefont{E.}~\bibnamefont{Nakamura}},
  \bibinfo{journal}{J. Phys. Soc. Japan} \textbf{\bibinfo{volume}{57}},
  \bibinfo{pages}{2836} (\bibinfo{year}{1988}).

\bibitem[{\citenamefont{Ehrhardt and Michel}(1981)}]{nano2:michel}
\bibinfo{author}{\bibfnamefont{K.~D.} \bibnamefont{Ehrhardt}} \bibnamefont{and}
  \bibinfo{author}{\bibfnamefont{K.~H.} \bibnamefont{Michel}},
  \bibinfo{journal}{Phys. Rev. Lett.} \textbf{\bibinfo{volume}{46}},
  \bibinfo{pages}{291} (\bibinfo{year}{1981}).

\bibitem[{\citenamefont{Kinase and Takahashi}(1992)}]{nano2:kinase}
\bibinfo{author}{\bibfnamefont{W.}~\bibnamefont{Kinase}} \bibnamefont{and}
  \bibinfo{author}{\bibfnamefont{K.}~\bibnamefont{Takahashi}},
  \bibinfo{journal}{J. Phys. Soc. Japan} \textbf{\bibinfo{volume}{61}},
  \bibinfo{pages}{329} (\bibinfo{year}{1992}).

\bibitem[{\citenamefont{Klein et~al.}(1982)\citenamefont{Klein, McDonald, and
  Ozaki}}]{nano2:klein}
\bibinfo{author}{\bibfnamefont{M.~L.} \bibnamefont{Klein}},
  \bibinfo{author}{\bibfnamefont{I.~R.} \bibnamefont{McDonald}},
  \bibnamefont{and} \bibinfo{author}{\bibfnamefont{Y.}~\bibnamefont{Ozaki}},
  \bibinfo{journal}{Phys.\ Rev.\ Lett.} \textbf{\bibinfo{volume}{48}},
  \bibinfo{pages}{1197} (\bibinfo{year}{1982}).

\bibitem[{\citenamefont{Klein and McDonald}(1982)}]{nano2:klein:london}
\bibinfo{author}{\bibfnamefont{M.~L.} \bibnamefont{Klein}} \bibnamefont{and}
  \bibinfo{author}{\bibfnamefont{I.~R.} \bibnamefont{McDonald}},
  \bibinfo{journal}{Proc. R. Soc. Lond.} \textbf{\bibinfo{volume}{A382}},
  \bibinfo{pages}{471} (\bibinfo{year}{1982}).

\bibitem[{\citenamefont{Lynden-bell et~al.}(1986)\citenamefont{Lynden-bell,
  Impey, and Klein}}]{nano2:klein_lynden-bell}
\bibinfo{author}{\bibfnamefont{R.~M.} \bibnamefont{Lynden-bell}},
  \bibinfo{author}{\bibfnamefont{R.~W.} \bibnamefont{Impey}}, \bibnamefont{and}
  \bibinfo{author}{\bibfnamefont{M.~L.} \bibnamefont{Klein}},
  \bibinfo{journal}{Chem.\ Phys.} \textbf{\bibinfo{volume}{109}},
  \bibinfo{pages}{25} (\bibinfo{year}{1986}).

\bibitem[{\citenamefont{Lu and Hardy}(1993)}]{hardy:nano2}
\bibinfo{author}{\bibfnamefont{H.~M.} \bibnamefont{Lu}} \bibnamefont{and}
  \bibinfo{author}{\bibfnamefont{J.~R.} \bibnamefont{Hardy}},
  \bibinfo{journal}{Solid State Commun.} \textbf{\bibinfo{volume}{87}},
  \bibinfo{pages}{1151} (\bibinfo{year}{1993}).

\bibitem[{\citenamefont{Lu and Hardy}(1990{\natexlab{a}})}]{hardy:k2seo4}
\bibinfo{author}{\bibfnamefont{H.~M.} \bibnamefont{Lu}} \bibnamefont{and}
  \bibinfo{author}{\bibfnamefont{J.~R.} \bibnamefont{Hardy}},
  \bibinfo{journal}{Phys.\ Rev.\ Lett.} \textbf{\bibinfo{volume}{64}},
  \bibinfo{pages}{661} (\bibinfo{year}{1990}{\natexlab{a}}).

\bibitem[{\citenamefont{Edwardson et~al.}(1989)\citenamefont{Edwardson, Boyer,
  Newman, Fox, Hardy, Flocken, Guenther, and Mei}}]{hardy:nacaf3}
\bibinfo{author}{\bibfnamefont{P.~J.} \bibnamefont{Edwardson}},
  \bibinfo{author}{\bibfnamefont{L.~L.} \bibnamefont{Boyer}},
  \bibinfo{author}{\bibfnamefont{R.~L.} \bibnamefont{Newman}},
  \bibinfo{author}{\bibfnamefont{D.~H.} \bibnamefont{Fox}},
  \bibinfo{author}{\bibfnamefont{J.~R.} \bibnamefont{Hardy}},
  \bibinfo{author}{\bibfnamefont{J.~W.} \bibnamefont{Flocken}},
  \bibinfo{author}{\bibfnamefont{R.~A.} \bibnamefont{Guenther}},
  \bibnamefont{and} \bibinfo{author}{\bibfnamefont{W.}~\bibnamefont{Mei}},
  \bibinfo{journal}{Phys.\ Rev.\ B} \textbf{\bibinfo{volume}{39}},
  \bibinfo{pages}{9738} (\bibinfo{year}{1989}).

\bibitem[{\citenamefont{Lu and Hardy}(1990{\natexlab{b}})}]{hardy:k2seo4_PRB}
\bibinfo{author}{\bibfnamefont{H.~M.} \bibnamefont{Lu}} \bibnamefont{and}
  \bibinfo{author}{\bibfnamefont{J.~R.} \bibnamefont{Hardy}},
  \bibinfo{journal}{Phys.\ Rev.\ B} \textbf{\bibinfo{volume}{42}},
  \bibinfo{pages}{8339} (\bibinfo{year}{1990}{\natexlab{b}}).

\bibitem[{\citenamefont{Liu et~al.}(2002{\natexlab{a}})\citenamefont{Liu, Duan,
  Mei, Smith, and Hardy}}]{hardy:clo4}
\bibinfo{author}{\bibfnamefont{J.}~\bibnamefont{Liu}},
  \bibinfo{author}{\bibfnamefont{C.}~\bibnamefont{Duan}},
  \bibinfo{author}{\bibfnamefont{W.~N.} \bibnamefont{Mei}},
  \bibinfo{author}{\bibfnamefont{R.~W.} \bibnamefont{Smith}}, \bibnamefont{and}
  \bibinfo{author}{\bibfnamefont{J.~R.} \bibnamefont{Hardy}},
  \bibinfo{journal}{J. Solid State Chem.} \textbf{\bibinfo{volume}{163}},
  \bibinfo{pages}{294} (\bibinfo{year}{2002}{\natexlab{a}}).

\bibitem[{\citenamefont{Liu et~al.}(1991)\citenamefont{Liu, Lu, Ullman, and
  Hardy}}]{hardy:k2so4}
\bibinfo{author}{\bibfnamefont{D.}~\bibnamefont{Liu}},
  \bibinfo{author}{\bibfnamefont{H.~M.} \bibnamefont{Lu}},
  \bibinfo{author}{\bibfnamefont{F.~G.} \bibnamefont{Ullman}},
  \bibnamefont{and} \bibinfo{author}{\bibfnamefont{J.~R.} \bibnamefont{Hardy}},
  \bibinfo{journal}{Phys.\ Rev.\ B} \textbf{\bibinfo{volume}{43}},
  \bibinfo{pages}{6202} (\bibinfo{year}{1991}).

\bibitem[{\citenamefont{Liu et~al.}(2002{\natexlab{b}})\citenamefont{Liu, Duan,
  Mei, Smith, and Hardy}}]{hardy:ca2sio4-sr2sio4}
\bibinfo{author}{\bibfnamefont{J.}~\bibnamefont{Liu}},
  \bibinfo{author}{\bibfnamefont{C.}~\bibnamefont{Duan}},
  \bibinfo{author}{\bibfnamefont{W.~N.} \bibnamefont{Mei}},
  \bibinfo{author}{\bibfnamefont{R.~W.} \bibnamefont{Smith}}, \bibnamefont{and}
  \bibinfo{author}{\bibfnamefont{J.~R.} \bibnamefont{Hardy}},
  \bibinfo{journal}{J. Chem. Phys.} \textbf{\bibinfo{volume}{116}},
  \bibinfo{pages}{3864} (\bibinfo{year}{2002}{\natexlab{b}}).

\bibitem[{\citenamefont{Lu and Hardy}(1991)}]{hardy:kno3}
\bibinfo{author}{\bibfnamefont{H.~M.} \bibnamefont{Lu}} \bibnamefont{and}
  \bibinfo{author}{\bibfnamefont{J.~R.} \bibnamefont{Hardy}},
  \bibinfo{journal}{Phys.\ Rev.\ B} \textbf{\bibinfo{volume}{44}},
  \bibinfo{pages}{7215} (\bibinfo{year}{1991}).

\bibitem[{\citenamefont{Liu et~al.}(2001)\citenamefont{Liu, Duan, Ossowski,
  Mei, Smith, and Hardy}}]{hardy:rbno3-csno3}
\bibinfo{author}{\bibfnamefont{J.}~\bibnamefont{Liu}},
  \bibinfo{author}{\bibfnamefont{C.}~\bibnamefont{Duan}},
  \bibinfo{author}{\bibfnamefont{M.~M.} \bibnamefont{Ossowski}},
  \bibinfo{author}{\bibfnamefont{W.~N.} \bibnamefont{Mei}},
  \bibinfo{author}{\bibfnamefont{R.~W.} \bibnamefont{Smith}}, \bibnamefont{and}
  \bibinfo{author}{\bibfnamefont{J.~R.} \bibnamefont{Hardy}},
  \bibinfo{journal}{J. Solid State Chem.} \textbf{\bibinfo{volume}{160}},
  \bibinfo{pages}{222} (\bibinfo{year}{2001}).

\bibitem[{\citenamefont{Liu et~al.}(2002{\natexlab{c}})\citenamefont{Liu, Duan,
  Ossowski, Mei, Smith, and Hardy}}]{hardy:agno3}
\bibinfo{author}{\bibfnamefont{J.}~\bibnamefont{Liu}},
  \bibinfo{author}{\bibfnamefont{C.}~\bibnamefont{Duan}},
  \bibinfo{author}{\bibfnamefont{M.~M.} \bibnamefont{Ossowski}},
  \bibinfo{author}{\bibfnamefont{W.~N.} \bibnamefont{Mei}},
  \bibinfo{author}{\bibfnamefont{R.~W.} \bibnamefont{Smith}}, \bibnamefont{and}
  \bibinfo{author}{\bibfnamefont{J.~R.} \bibnamefont{Hardy}},
  \bibinfo{journal}{J. Phys. Chem. Solids} \textbf{\bibinfo{volume}{63}},
  \bibinfo{pages}{409} (\bibinfo{year}{2002}{\natexlab{c}}).

\bibitem[{\citenamefont{Hehre et~al.}(1986)\citenamefont{Hehre, Radom,
  Schleyer, and Pople}}]{abinitio:hehre}
\bibinfo{author}{\bibfnamefont{W.}~\bibnamefont{Hehre}},
  \bibinfo{author}{\bibfnamefont{L.}~\bibnamefont{Radom}},
  \bibinfo{author}{\bibfnamefont{P.}~\bibnamefont{Schleyer}}, \bibnamefont{and}
  \bibinfo{author}{\bibfnamefont{J.}~\bibnamefont{Pople}},
  \emph{\bibinfo{title}{Ab initio molecular orbital theory}}
  (\bibinfo{publisher}{John Wiley \& Sons}, \bibinfo{address}{New York},
  \bibinfo{year}{1986}), \bibinfo{note}{p. 336}.

\bibitem[{\citenamefont{Ravindran et~al.}(1999)\citenamefont{Ravindran, Delin,
  Johansson, Eriksson, and Wills}}]{nano2:ravindran}
\bibinfo{author}{\bibfnamefont{P.}~\bibnamefont{Ravindran}},
  \bibinfo{author}{\bibfnamefont{A.}~\bibnamefont{Delin}},
  \bibinfo{author}{\bibfnamefont{B.}~\bibnamefont{Johansson}},
  \bibinfo{author}{\bibfnamefont{O.}~\bibnamefont{Eriksson}}, \bibnamefont{and}
  \bibinfo{author}{\bibfnamefont{J.~M.} \bibnamefont{Wills}},
  \bibinfo{journal}{Phys. Rev. B} \textbf{\bibinfo{volume}{59}},
  \bibinfo{pages}{1776} (\bibinfo{year}{1999}).

\bibitem[{\citenamefont{Fowler and Madden}(1984)}]{alkali_halide:fowler}
\bibinfo{author}{\bibfnamefont{P.~W.} \bibnamefont{Fowler}} \bibnamefont{and}
  \bibinfo{author}{\bibfnamefont{P.~A.} \bibnamefont{Madden}},
  \bibinfo{journal}{Phys. Rev. B} \textbf{\bibinfo{volume}{29}},
  \bibinfo{pages}{1035} (\bibinfo{year}{1984}).

\bibitem[{\citenamefont{Kim and
  Gordon}(1974)}]{gordon:kim:alkali-halide_alkaline-earth-dihalide}
\bibinfo{author}{\bibfnamefont{Y.~S.} \bibnamefont{Kim}} \bibnamefont{and}
  \bibinfo{author}{\bibfnamefont{R.~G.} \bibnamefont{Gordon}},
  \bibinfo{journal}{Phys. Rev. B} \textbf{\bibinfo{volume}{9}},
  \bibinfo{pages}{3548} (\bibinfo{year}{1974}).

\bibitem[{\citenamefont{Cohen and Gordon}(1975)}]{gordon:cohen:alkali-halide}
\bibinfo{author}{\bibfnamefont{A.~J.} \bibnamefont{Cohen}} \bibnamefont{and}
  \bibinfo{author}{\bibfnamefont{R.~G.} \bibnamefont{Gordon}},
  \bibinfo{journal}{Phys.\ Rev.\ B} \textbf{\bibinfo{volume}{12}},
  \bibinfo{pages}{3228} (\bibinfo{year}{1975}).

\bibitem[{\citenamefont{Boyer}(1979)}]{boyer:nacl-kcl}
\bibinfo{author}{\bibfnamefont{L.~L.} \bibnamefont{Boyer}},
  \bibinfo{journal}{Phys.\ Rev.\ Lett.} \textbf{\bibinfo{volume}{42}},
  \bibinfo{pages}{584} (\bibinfo{year}{1979}).

\bibitem[{\citenamefont{Sepliarsky et~al.}(2000)\citenamefont{Sepliarsky,
  Phillpot, Wolf, Stachiotti, and Migoni}}]{phillpot}
\bibinfo{author}{\bibfnamefont{M.}~\bibnamefont{Sepliarsky}},
  \bibinfo{author}{\bibfnamefont{S.~R.} \bibnamefont{Phillpot}},
  \bibinfo{author}{\bibfnamefont{D.}~\bibnamefont{Wolf}},
  \bibinfo{author}{\bibfnamefont{M.~G.} \bibnamefont{Stachiotti}},
  \bibnamefont{and} \bibinfo{author}{\bibfnamefont{R.~L.}
  \bibnamefont{Migoni}}, \bibinfo{journal}{Appl.\ Phys.\ Lett.}
  \textbf{\bibinfo{volume}{76}}, \bibinfo{pages}{3986} (\bibinfo{year}{2000}).

\bibitem[{\citenamefont{{\it et al.}}(2002)}]{gaussian}
\bibinfo{author}{\bibfnamefont{M.~J.~F.} \bibnamefont{{\it et al.}}},
  \emph{\bibinfo{title}{GAUSSIAN 98, Revision A.11.3}}
  (\bibinfo{publisher}{Gaussian, Inc.}, \bibinfo{address}{Pittsburgh PA},
  \bibinfo{year}{2002}).

\bibitem[{\citenamefont{Gordon and Kim}(1972)}]{gordon:kim}
\bibinfo{author}{\bibfnamefont{R.~G.} \bibnamefont{Gordon}} \bibnamefont{and}
  \bibinfo{author}{\bibfnamefont{Y.~S.} \bibnamefont{Kim}},
  \bibinfo{journal}{J.\ Chem.\ Phys.} \textbf{\bibinfo{volume}{56}},
  \bibinfo{pages}{3122} (\bibinfo{year}{1972}).

\bibitem[{\citenamefont{Waldman and
  Gordon}(1979{\natexlab{a}})}]{gordon:waldman:shell}
\bibinfo{author}{\bibfnamefont{M.}~\bibnamefont{Waldman}} \bibnamefont{and}
  \bibinfo{author}{\bibfnamefont{R.~G.} \bibnamefont{Gordon}},
  \bibinfo{journal}{J. Chem. Phys.} \textbf{\bibinfo{volume}{71}},
  \bibinfo{pages}{1340} (\bibinfo{year}{1979}{\natexlab{a}}).

\bibitem[{\citenamefont{Duan et~al.}(2001)\citenamefont{Duan, Mei, Smith, Liu,
  Ossowski, and Hardy}}]{hardy:nitrite}
\bibinfo{author}{\bibfnamefont{C.~G.} \bibnamefont{Duan}},
  \bibinfo{author}{\bibfnamefont{W.~N.} \bibnamefont{Mei}},
  \bibinfo{author}{\bibfnamefont{R.~W.} \bibnamefont{Smith}},
  \bibinfo{author}{\bibfnamefont{J.}~\bibnamefont{Liu}},
  \bibinfo{author}{\bibfnamefont{M.~M.} \bibnamefont{Ossowski}},
  \bibnamefont{and} \bibinfo{author}{\bibfnamefont{J.~R.} \bibnamefont{Hardy}},
  \bibinfo{journal}{Phys.\ Rev.\ B} \textbf{\bibinfo{volume}{63}},
  \bibinfo{pages}{144105} (\bibinfo{year}{2001}).

\bibitem[{\citenamefont{Parker et~al.}(1975)\citenamefont{Parker, Snow, and
  Pack}}]{hf:parker}
\bibinfo{author}{\bibfnamefont{G.~A.} \bibnamefont{Parker}},
  \bibinfo{author}{\bibfnamefont{R.~L.} \bibnamefont{Snow}}, \bibnamefont{and}
  \bibinfo{author}{\bibfnamefont{R.~T.} \bibnamefont{Pack}},
  \bibinfo{journal}{Chem. Phys. Lett.} \textbf{\bibinfo{volume}{33}},
  \bibinfo{pages}{399} (\bibinfo{year}{1975}).

\bibitem[{\citenamefont{Banerjee et~al.}(1980)\citenamefont{Banerjee, Shepard,
  and Simons}}]{no2-h2o:banerjee}
\bibinfo{author}{\bibfnamefont{A.}~\bibnamefont{Banerjee}},
  \bibinfo{author}{\bibfnamefont{R.}~\bibnamefont{Shepard}}, \bibnamefont{and}
  \bibinfo{author}{\bibfnamefont{J.}~\bibnamefont{Simons}},
  \bibinfo{journal}{J. Chem. Phys.} \textbf{\bibinfo{volume}{73}},
  \bibinfo{pages}{1814} (\bibinfo{year}{1980}).

\bibitem[{\citenamefont{Clementi and Roetti}(1974)}]{clementi}
\bibinfo{author}{\bibfnamefont{E.}~\bibnamefont{Clementi}} \bibnamefont{and}
  \bibinfo{author}{\bibfnamefont{C.}~\bibnamefont{Roetti}},
  \bibinfo{journal}{At. Data Nucl. Data Tables} \textbf{\bibinfo{volume}{14}},
  \bibinfo{pages}{177} (\bibinfo{year}{1974}).

\bibitem[{\citenamefont{Kay}(1972)}]{nano2:kay}
\bibinfo{author}{\bibfnamefont{M.~I.} \bibnamefont{Kay}},
  \bibinfo{journal}{Ferroelectrics} \textbf{\bibinfo{volume}{4}},
  \bibinfo{pages}{235} (\bibinfo{year}{1972}).

\bibitem[{\citenamefont{Hartwig et~al.}(1972)\citenamefont{Hartwig,
  Wiener-Avnear, and Porto}}]{nano2:hartwig}
\bibinfo{author}{\bibfnamefont{C.~M.} \bibnamefont{Hartwig}},
  \bibinfo{author}{\bibfnamefont{E.}~\bibnamefont{Wiener-Avnear}},
  \bibnamefont{and} \bibinfo{author}{\bibfnamefont{S.~P.~S.}
  \bibnamefont{Porto}}, \bibinfo{journal}{Phys. Rev. B}
  \textbf{\bibinfo{volume}{5}}, \bibinfo{pages}{79} (\bibinfo{year}{1972}).

\bibitem[{\citenamefont{Okuda et~al.}(1990)\citenamefont{Okuda, Ohba, Saito,
  Ito, and Shibuya}}]{nano2:okuda}
\bibinfo{author}{\bibfnamefont{M.}~\bibnamefont{Okuda}},
  \bibinfo{author}{\bibfnamefont{S.}~\bibnamefont{Ohba}},
  \bibinfo{author}{\bibfnamefont{Y.}~\bibnamefont{Saito}},
  \bibinfo{author}{\bibfnamefont{T.}~\bibnamefont{Ito}}, \bibnamefont{and}
  \bibinfo{author}{\bibfnamefont{I.}~\bibnamefont{Shibuya}},
  \bibinfo{journal}{Acta Cryst. B} \textbf{\bibinfo{volume}{46}},
  \bibinfo{pages}{343} (\bibinfo{year}{1990}).

\bibitem[{\citenamefont{Parrinello and Rahman}(1980)}]{parrinello-rahman:80}
\bibinfo{author}{\bibfnamefont{M.}~\bibnamefont{Parrinello}} \bibnamefont{and}
  \bibinfo{author}{\bibfnamefont{A.}~\bibnamefont{Rahman}},
  \bibinfo{journal}{Phys.\ Rev.\ Lett.} \textbf{\bibinfo{volume}{45}},
  \bibinfo{pages}{1196} (\bibinfo{year}{1980}).

\bibitem[{\citenamefont{Waldman and
  Gordon}(1979{\natexlab{b}})}]{gordon:waldman:scale}
\bibinfo{author}{\bibfnamefont{M.}~\bibnamefont{Waldman}} \bibnamefont{and}
  \bibinfo{author}{\bibfnamefont{R.~G.} \bibnamefont{Gordon}},
  \bibinfo{journal}{J.\ Chem.\ Phys.} \textbf{\bibinfo{volume}{71}},
  \bibinfo{pages}{1325} (\bibinfo{year}{1979}{\natexlab{b}}).

\bibitem[{\citenamefont{Fowler and Klein}(1986)}]{nacn:fowler}
\bibinfo{author}{\bibfnamefont{P.~W.} \bibnamefont{Fowler}} \bibnamefont{and}
  \bibinfo{author}{\bibfnamefont{M.~L.} \bibnamefont{Klein}},
  \bibinfo{journal}{J.\ Chem.\ Phys.} \textbf{\bibinfo{volume}{85}},
  \bibinfo{pages}{3913} (\bibinfo{year}{1986}).

\bibitem[{\citenamefont{Kremer and Siems}(1988)}]{nano2:kremer}
\bibinfo{author}{\bibfnamefont{J.~W.} \bibnamefont{Kremer}} \bibnamefont{and}
  \bibinfo{author}{\bibfnamefont{R.}~\bibnamefont{Siems}},
  \bibinfo{journal}{Ferroelectrics} \textbf{\bibinfo{volume}{79}},
  \bibinfo{pages}{35} (\bibinfo{year}{1988}).

\end{thebibliography}
\end{document}